\documentclass[aps,prb,twocolumn,superscriptaddress]{revtex4-1}
\usepackage{longtable}
\usepackage{morefloats}
\usepackage[dvips]{graphicx}
\usepackage{color}
\usepackage{epsfig,graphicx,amsfonts,amsbsy}
\usepackage{amsmath,amsfonts,amsthm,amssymb}
\usepackage{appendix}
\usepackage{makeidx}
\usepackage{url}
\usepackage[bookmarksnumbered,pdfpagelabels=true,plainpages=false,colorlinks=true,linkcolor=blue,citecolor=green,urlcolor=blue]{hyperref}

\begin{document}
\title{Spin-wave spectra in periodically surface-modulated ferromagnetic thin films}
\author{R. A. Gallardo}
\affiliation{Departamento de F\'{i}sica, Universidad T\'{e}cnica Federico Santa Mar\'{i}a, Avenida Espa\~{n}a 1680, Valpara\'{i}so, Chile}
\author{M. Langer}
\affiliation{Institute of Ion Beam Physics and Materials Research, Helmholtz-Zentrum Dresden-Rossendorf e.V., P.O. Box 510119, 01314 Dresden, Germany}
\author{A. Rold\'an-Molina}
\affiliation{Instituto de F\'{i}sica, Pontificia Universidad Cat\'{o}lica de Valpara\'{i}so, Avenida Brasil 2950, Valpara\'{i}so, Chile}
\author{T. Schneider}
\affiliation{Institute of Ion Beam Physics and Materials Research, Helmholtz-Zentrum Dresden-Rossendorf e.V., P.O. Box 510119, 01314 Dresden, Germany}
\author{K. Lenz}
\affiliation{Institute of Ion Beam Physics and Materials Research, Helmholtz-Zentrum Dresden-Rossendorf e.V., P.O. Box 510119, 01314 Dresden, Germany}
\author{J. Lindner}
\affiliation{Institute of Ion Beam Physics and Materials Research, Helmholtz-Zentrum Dresden-Rossendorf e.V., P.O. Box 510119, 01314 Dresden, Germany}
\author{P. Landeros}
\affiliation{Departamento de F\'{i}sica, Universidad T\'{e}cnica Federico Santa Mar\'{i}a, Avenida Espa\~{n}a 1680, Valpara\'{i}so, Chile}
\date{\today }

%%%%%%%%%%%%%%ABSTRACT%%%%%%%%%%%%%%
\begin{abstract}

This article presents theoretical results for the dynamic response of periodically surface-modulated ferromagnetic thin films. 
For such system, the role of the periodic dipolar field induced by the modulation is addressed by using the plane-wave method. 
By controlling the geometry of the modulated volumes within the film, the frequency modes and spatial profiles of spin waves can be manipulated.
The angular dependence of the frequency band-gaps unveils the influence of both dynamic and static magnetic charges, which reside in the edges of the etching periodic zones, and it is stablished that band-gap widths created by static magnetic charges are broader than the one created by dynamic magnetic charges.
To corroborate the validity of the model, the theoretical results are compared with ferromagnetic resonance simulations, where a very good agreement is achieved between both methods.
The theoretical model allows for a detailed understanding of the physics underlying these kind of systems, thereby providing an outlook to potential applications associated with magnonic crystals-based devices.

\end{abstract}
%%%%%%%%%%%%%%ABSTRACT%%%%%%%%%%%%%%
\maketitle

%%%%%%%%%%%%%%INTRODUCCTION%%%%%%%%%%%%%%
\section{Introduction}
\label{SI}

Spin waves (SWs) at microwave frequencies are of current potential interest for wireless communications technologies, since they can carry and handle information in a unique way.\cite{Demokritov12,Chumak15}
Such waves are able to carry pure spin currents (currents without charge transport which can be then properly converted into measurable charge currents), even in magnetic insulators.\cite{Kajiwara10}
This relevant property of magnon based technologies has the key advantage of substantially reducing the energy waste due to Joule heating, one of the main drawbacks inherent of conventional electronics. 
In this way, spin waves provide a new way to exploit the collective behavior of the electrons in a solid.  
SWs have further been proposed as building blocks for computational architectures allowing to perform logic operations.\cite{Khitun05,Khitun10,Jamali13} 
One of the potential aspects of spin-wave based technologies is that both the amplitude and the phase of spin waves may encode information.\cite{Kostylev05,Tacchi15} 
Besides, the non-linearity of the spin waves permits the realization of a magnon transistor, whose basic principle relies on four-magnon scattering processes.\cite{Chumak14} 

Manipulating spin-wave propagation by means of periodic modulation of the magnetic properties within thin films nowadays can be regarded as an important research field in magnetism.\cite{Vasseur96,Nikitov01,Puszkarski03,Kruglyak06,Gubbiotti07,Krawczyk08,Neusser08,Chumak09,Lee09,Neusser09,Wang09,Kim09,Serga10,Kruglyak10,Wang10,Gubbiotti10,Cao10,Ding11,Lenk11,Tacchi11,Tacchi12,Chumak12,Yu13,Krawczyk14,Korner13,Gallardo14,Sebastian15}
Such research area is currently named magnonics or magnon-spintronics, and it is based on the control of spin waves in periodic magnetic structures called magnonic crystals (MCs).\cite{Demokritov12,Chumak14,Chumak15} 
The possibility of such system to act as a spin wave filter with a pronounced discretization of the SWs frequency turns out key for applications in signal processing and storage-recovery mechanisms.\cite{Kim09,Chumak12}
In this context, MCs have been extensively studied, since they exhibit adjustable frequency band gaps (BGs), which can be optimized by modulating the magnetic parameters or changing the geometry and arrangement of periodic scattering centers.\cite{Vasseur96,Nikitov01,Puszkarski03,Kruglyak06,Gubbiotti07,Krawczyk08,Neusser08,Chumak09,Lee09,Neusser09,Wang09,Kim09,Serga10,Kruglyak10,Wang10,Gubbiotti10,Cao10,Ding11,Lenk11,Tacchi11,Tacchi12,Chumak12,Yu13,Krawczyk14,Tacchi15,Sebastian15}
The design of the MCs can be realized by artificial modulation of the magnetic properties, either in the form of dipolarly coupled nanowires\cite{Gubbiotti07}, bicomponent magnonic crystals,\cite{Wang09,Wang10,Tacchi12} width-modulated waveguides,\cite{Chumak09,Lee09,Kim09,Ding11,Ciubotaru12} antidot lattices,\cite{Neusser08,Gubbiotti10,Lenk11,Klos12,Krawczyk13} step-modulated thickness nanowires,\cite{Gubbiotti14} or by means of ion-implantation.\cite{Barsukov11,Obry13,Landeros12,Gallardo14}
Furthermore, dynamic magnonic crystals have also been investigated, where the periodic magnetic field, for instance, originates from a meander-like current-carrying wire,\cite{Chumak09b,Chumak14b} or even by locally heating the magnetic material.\cite{Vogel15,Grundler15}

A large variety of studies based on Brillouin light scattering (BLS) have been carried out on magnonic crystals, where the presence of frequency band gaps has been confirmed and accomplished with theoretical results.\cite{Gubbiotti07,Chumak09,Wang09,Wang10,Gubbiotti10,Ding11,Lenk11,Tacchi11,Tacchi12}
Moreover, bi-component MCs have been studied,\cite{Krawczyk08,Wang09,Lin11,Sokolovskyy11,Mruczkiewicz13} where periodic properties originate from a different saturation magnetization $M_s$, anisotropy $K$ or exchange constant $A$. 
Thus, modification of these parameters allows for controlling the BG position and the localization of SWs.
For instance, increasing the difference of $M_s$ of a bi-component MC can lead to a broadening of the BG frequency range and enable the concentration of a spin-wave excitation within the zone of lower or higher saturation magnetization.
However, experimentally defining material parameters such as magnetization or exchange length with laterally well-defined periodicities often is not straightforward and suffering from limitations of the range in which variations are possible for a given material.
Therefore, a periodic geometrical modulation is an interesting alternative to create a kind of magnonic crystal, where the role of the contrast between material can be replaced by the size of the periodic modulation of the surface.

In this paper, a periodically surface modulated ferromagnetic thin film is studied theoretically, as described in Sec. \ref{SII}, and corroborated with results obtained by micromagnetic simulations. 
The theory is based on the plane-wave method (PWM) (see e.g. Refs. \onlinecite{Krawczyk08} and  \onlinecite{Sokolovskyy11}) and the small wave vector limit is directly compared with the numerical simulations.
The discussion of the results is presented in Sec. \ref{SIII}, while the final conclusions are highlighted in \ref{SIV}.  
%%%%%%%%%%%%%%INTRODUCCTION%%%%%%%%%%%%%%

%%%%%%%%%%%%%%THEORY%%%%%%%%%%%%%%
\section{Theoretical description}
\label{SII}
\begin{figure}[h]
\includegraphics[width=8.5 cm]{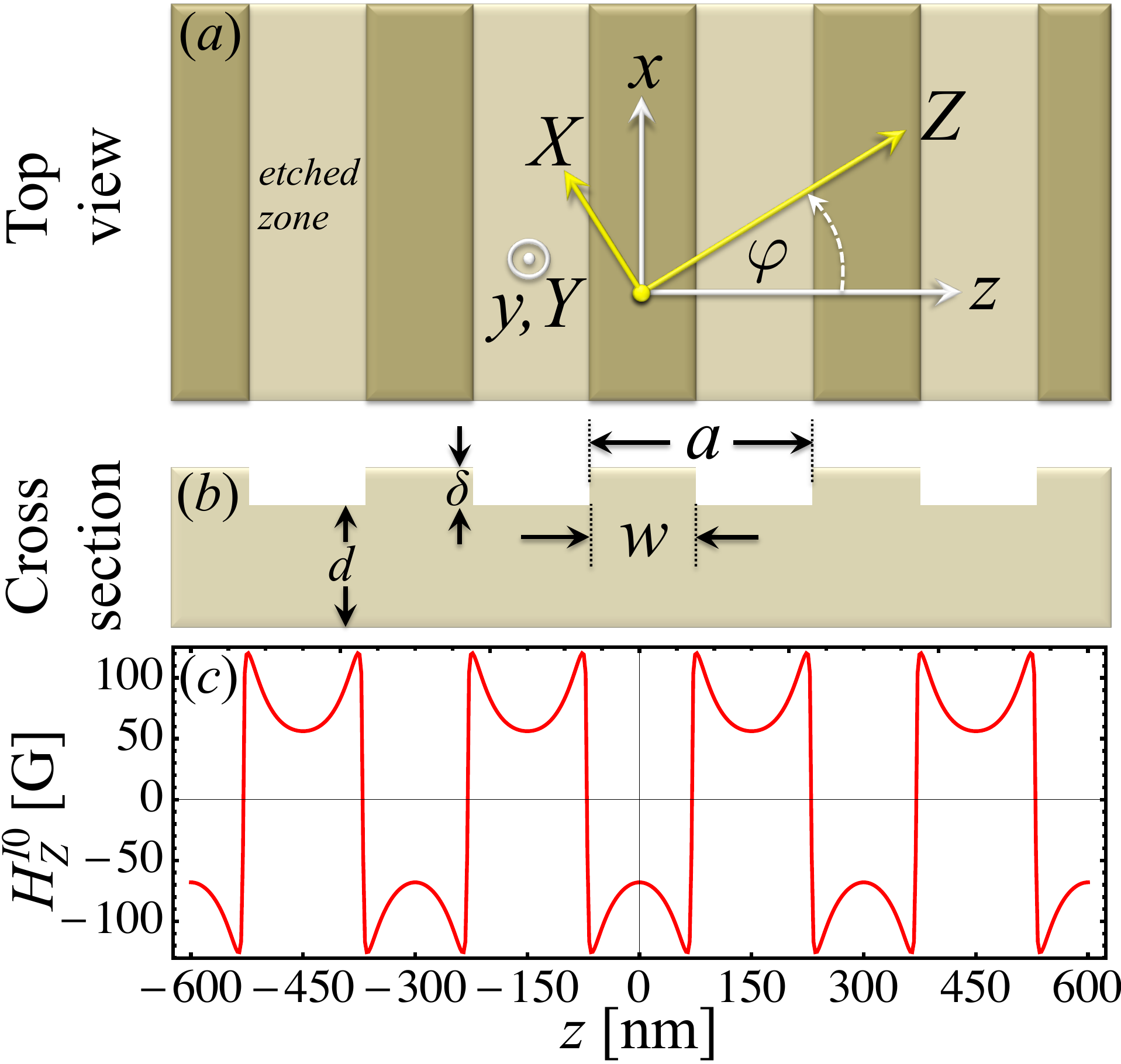}
\caption{In ($a$) the top view is shown, where two reference systems are depicted.
The coordinates ($x$,$y$,$z$) are defined by the periodic structure, while ($X$,$Y$,$Z$) is defined according to the equilibrium orientation of the magnetization, which points along $Z$. 
Because the magnetization lies in the film's plane, $y$ and $Y$ are matching.
In (b) the cross section is depicted, where the main geometrical parameters are defined.
Figure (c) shows the periodic dipolar field [see Eq. (\ref{H(0)Z})] created by the static magnetic charges at the edges of the etched zone for $\delta=2$ nm.
Additional parameters are given in section \ref{SIII}.}
\label{Fig1}
\end{figure}

In bi-component magnonic crystals,\cite{Wang09,Wang10,Tacchi12} the periodic properties originate from the contrast between different ferromagnetic materials with different magnetic parameters, for instance the saturation magnetization $M_s$ or the micromagnetic exchange constant $A$. 
Nevertheless in surface-modulated magnonic crystals the periodic properties arise from the periodic magnetic charges created at the edges of the etched zones, as shown Fig. \ref{Fig1}($b$) for a one-dimensional surface-modulated thin film.
%---------------FIGURE 1---------------
%---------------FIGURE 1---------------

The temporal evolution of the system can be described using the Landau-Lifshitz (LL) equation $\mathbf{\dot{M}}(\mathbf{r};t)=-\gamma \mathbf{M}(\mathbf{r};t)\times\mathbf{H}^{e}(\mathbf{r};t)$. 
Here, $\gamma$ is the absolute value of the gyromagnetic ratio, $\mathbf{M}(\mathbf{r};t)$ is the magnetization and $\mathbf{H}^{e}(\mathbf{r};t)$ is the effective field.
For small deviations around the equilibrium, both the magnetization and the effective field are written as $\mathbf{M}(\mathbf{r};t)=M_s\hat{Z}+\mathbf{m}(\mathbf{r};t)$ and $\mathbf{H}^{e}(\mathbf{r};t)=\mathbf{H}^{e0}(\mathbf{r})+\mathbf{h}^e(\mathbf{r};t)$, respectively.
Note that $\hat{Z}$ points along the equilibrium orientation of the magnetization, which lies in-plane and $\mathbf{h}^e(\mathbf{r};t)$ is proportional to the dynamic magnetization $\mathbf{m}(\mathbf{r};t)$.
Thus, in the linear regime, the LL equation reads
\begin{subequations}
\begin{eqnarray}
i\Omega m_X(\mathbf{r})&=&-m_Y(\mathbf{r})H^{e0}_Z(\mathbf{r}) +M_sh^{e}_Y(\mathbf{r})
\\
i\Omega m_Y( \mathbf{r})&=&m_X(\mathbf{r})H^{e0}_Z(\mathbf{r}) -M_sh^{e}_X(\mathbf{r})
\end{eqnarray}
\label{LL1}
\end{subequations}
wherein it has been assumed $\mathbf{h}^e(\mathbf{r};t)=\mathbf{h}^e(\mathbf{r})e^{i\omega t}$, and then $\mathbf{m}(\mathbf{r};t)=\mathbf{m}(\mathbf{r})e^{i\omega t}$,  and we have also defined $\Omega=\omega/\gamma$.
Moreover,  note that $H^{e0}_{\eta}$ ($h^{e}_{\eta}$) is the $\eta$-component of the static (dynamic) effective magnetic field.
The effective field is defined as $\mathbf{H}^{e}(\mathbf{r})=\mathbf{H}+\mathbf{H}^{ex}(\mathbf{r})+\mathbf{H}^d(\mathbf{r})+\mathbf{H}^{I}(\mathbf{r})$, where $\mathbf{H}$ is the external field, $\mathbf{H}^{ex}(\mathbf{r})=(D_{ex}/M_s)\nabla^2\mathbf{M}(\mathbf{r})$ is the exchange field with $ D_{ex}(\mathbf{r})=2A/\mu_0 M_s$, wherein $A$ is the exchange stiffness constant.
Furthermore, $\mathbf{H}^d(\mathbf{r})$ is the dipolar field of the flat film and $\mathbf{H}^{I}(\mathbf{r})$ is the dipolar field induced by the periodic magnetic charges, which reside at the edges of the etched zones.
According to Fig. \ref{Fig1}, the periodic distribution of the etched regions of thickness $\delta$ over the top surface of the ferromagnetic film induces a periodic stray field that interacts with the magnetization of the nominal film of thickness $d$. 
In this way, according to Bloch's theorem, the dynamic components of the magnetization can be expanded into Fourier series as $\mathbf{m}(\mathbf{r})=\sum_{\mathbf{G}}\mathbf{m}(\mathbf{G})e^{i(\mathbf{G}+\mathbf{k})\cdot\mathbf{r}}$, where $\mathbf{G}=G_q\hat{x}+G_n\hat{z}$ denotes a reciprocal lattice vector, with $G_q=(2\pi/a_x)q$, $G_n=(2\pi/a_z) n$ and both $n$ and $q$ are integer numbers. 
The above picture considers a general two-dimensional periodic modulation of the etched zones, which can be easily adapted to one-dimensional periodic structures by setting $G_q=0$, as depicted in Fig. \ref{Fig1}.  
Thus, the dynamic components of the dipolar field averaged over the film's thickness are
\begin{eqnarray}
h^{d}_{Y}(\mathbf{r})&=& -4\pi\sum_{\mathbf{G}}m_{Y}(\mathbf{G})\zeta(\mathbf{G})e^{i(\mathbf{G}+\mathbf{k})\cdot\mathbf{r}}
\label{hdY}
\end{eqnarray}
and
\begin{equation}
h^{d}_{X}(\mathbf{r})= 4\pi\sum_{\mathbf{G}}m_{X}(\mathbf{G})\xi(\mathbf{G},\mathbf{k})^2 \frac{\zeta(\mathbf{G})-1}{\vert \mathbf{G}+\mathbf{k}\vert^2}e^{i(\mathbf{G}+\mathbf{k})\cdot\mathbf{r}},
\label{hdX}
\end{equation}
where
\begin{eqnarray}
\zeta(\mathbf{G})&=&\frac{2\sinh[\vert \mathbf{G}+\mathbf{k}\vert d/2]e^{-\left\vert \mathbf{G}+\mathbf{k} \right\vert d/2}}{\vert \mathbf{G}+\mathbf{k}\vert d}
\label{ZZ}
\end{eqnarray}
and
\begin{eqnarray}
\xi(\mathbf{G},\mathbf{k})&=&(G_n+k_z)\sin\varphi-(G_q+k_x)\cos\varphi.
\label{Xi}
\end{eqnarray}
In previous works, the dynamic dipolar fields (\ref{hdY}) and (\ref{hdX}) are evaluated at the middle of the film thickness ($y=d/2$),\cite{Sokolovskyy11,Klos12,Krawczyk13} where $\zeta( \mathbf{G})$ reduces to $e^{-\left\vert \mathbf{G}+\mathbf{k} \right\vert d/2}$. 
Nevertheless, under this simplifying assumption, systematic deviations from the simulation cannot be removed. 
Likewise, the exchange dynamic field components are 
\begin{equation}
h^{ex}_{X,Y}(\mathbf{r})=- \frac{D_{ex}}{M_s}\sum_{\mathbf{G}}(\mathbf{G}+\mathbf{k})^2m_{X,Y}( \mathbf{G})e^{i(\mathbf{G}+\mathbf{k})\cdot\mathbf{r}}.
\label{hEX}
\end{equation}

In order to obtain the periodic static field $\mathbf{H}^{I 0}(\mathbf{r})$, it is noted that the static magnetization components in the range $d+\delta>y>d$, can be written as
\begin{eqnarray}
 M_{z} &=&M_s\cos \varphi\sum_{\mathbf{G}}C_{\mathbf{G}}(y)\exp \left[i\mathbf{G}\cdot \mathbf{r}\right] 
\end{eqnarray}
and
\begin{eqnarray}
M_{x}& =&M_{s}\sin \varphi\sum_{\mathbf{G}}C_{\mathbf{G}}(y) \exp \left[ i\mathbf{G}\cdot \mathbf{r}\right].
\label{M7}
\end{eqnarray}
Then, following Ref.\ \onlinecite{Gallardo14}, the magnetostatic potential is given by
\begin{eqnarray}
\phi(\mathbf{r}) &=&-i M_s \sum_{\mathbf{G}}\chi(\mathbf{G})\int C_{\mathbf{G}}(y^{\prime })\frac{e^{i\mathbf{G}\mathbf{\cdot r^{\prime }}}}{\left\vert \mathbf{r}-\mathbf{r}^{\prime }\right\vert }\mathrm{d}^{3}\mathbf{r}^{\prime },
\end{eqnarray}
where $\chi(\mathbf{G})=G_{n}\cos \varphi+G_{q}\sin \varphi$. Besides, note that $C_{\mathbf{G}}(y^{\prime })=0$ for $y^{\prime }>d+\delta$ and $y^{\prime }<d$.
Therefore, an analytical expression can be derived for the magnetostatic potential, which is
\begin{eqnarray}
\phi(\mathbf{r}) &=&i2\pi M_s\nonumber
\sum_{\mathbf{G}}C_{\mathbf{G}}
\chi(\mathbf{G})
\frac{e^{\vert \mathbf{G}\vert (y-d-\delta)}\left(1-e^{\vert \mathbf{G}\vert \delta}\right)}{\vert \mathbf{G}\vert^2}
e^{i\mathbf{G}\mathbf{\cdot r}}.\nonumber
\label{PP1}
\end{eqnarray}
Now, the components of the static field are
\begin{eqnarray}
H_{X}^{I 0}(\mathbf{r}) &=&-2\pi  M_s\sum_{\mathbf{G}}C_{\mathbf{G}}\chi(\mathbf{G})\xi(\mathbf{G},0) \eta(\mathbf{G})
e^{i\mathbf{G}\mathbf{\cdot r}},
\label{H(0)X}
\end{eqnarray}
\begin{eqnarray}
H_{Z}^{I 0}(\mathbf{r}) &=&-2\pi  M_s\sum_{\mathbf{G}}C_{\mathbf{G}}\chi(\mathbf{G})^2 \eta(\mathbf{G})
e^{i\mathbf{G}\mathbf{\cdot r}},
\label{H(0)Z}
\end{eqnarray}
and
\begin{eqnarray}
H_{Y}^{I 0}(\mathbf{r}) &=&i2\pi M_s
\sum_{\mathbf{G}}C_{\mathbf{G}}
\chi(\mathbf{G})\eta(\mathbf{G})
\vert \mathbf{G}\vert
e^{i\mathbf{G}\mathbf{\cdot r}}.
\label{H(0)Y}
\end{eqnarray}
Here, the following definition was used
\begin{eqnarray}
\eta(\mathbf{G})&=&\frac{e^{-\vert \mathbf{G}\vert (d+\delta)}}{\vert \mathbf{G}\vert^3 d}(e^{\vert \mathbf{G}\vert d}-1)(e^{\vert \mathbf{G}\vert \delta}-1).
\label{eta}
\end{eqnarray}

In expressions (\ref{H(0)X})--(\ref{H(0)Y}), an average over the nominal FM film thickness $d$ has been performed, in such a way that at $d=0$, the magnetostatic potential $\phi(\mathbf{r})=0$. 
On the other hand, the dynamic magnetization components in the etched part can be written as 
\begin{eqnarray}
m_{X,Y}(\mathbf{r})&=&\sum_{\mathbf{G},\mathbf{G}^{\prime}}m_{X,Y}(\mathbf{G})C_{\mathbf{G}^{\prime}}e^{i(\mathbf{G} +\mathbf{G}^{\prime}+\mathbf{k})\cdot \mathbf{r}},
\label{MM1}
\end{eqnarray}
where it is assumed that the dynamic magnetization is uniform along the thickness. 
This approximation is valid for small values of depth $\delta$, nevertheless, when $\delta$ increases the boundary conditions may produce a modulation of spin waves along the thickness and therefore Eq.\ (\ref{MM1}) is not valid anymore. 
By using the same procedure to derive Eqs. (\ref{H(0)X})--(\ref{H(0)Y}), the components of the dynamic dipolar field $\mathbf{h}^{I}(\mathbf{r})$ derived from (\ref{MM1}) and averaged over the nominal film are
\begin{eqnarray}
h^{I}_{Y}(\mathbf{r})&=&2\pi\sum_{\mathbf{G},\mathbf{G}^{\prime}}C_{\mathbf{G}^{\prime}}e^{i(\mathbf{G}+\mathbf{G}^{\prime}+\mathbf{k})\cdot \mathbf{r}}\left\{m_{Y}(\mathbf{G})
\eta(\mathbf{G}+\mathbf{G}^{\prime}+\mathbf{k})\right.\nonumber
\\
&-&\left.i m_{X}(\mathbf{G})\xi(\mathbf{G}+\mathbf{G}^{\prime},\mathbf{k})
\frac{\eta(\mathbf{G}+\mathbf{G}^{\prime}+\mathbf{k})}{\left\vert  \mathbf{G}+\mathbf{G}^{\prime}+\mathbf{k}\right\vert}\right\},
\label{hYY}
\end{eqnarray}
and
\begin{eqnarray}
h^{I}_{X}(\mathbf{r})&=&-2\pi\sum_{\mathbf{G},\mathbf{G}^{\prime}}C_{\mathbf{G}^{\prime}}e^{i(\mathbf{G}+\mathbf{G}^{\prime}+\mathbf{k})\cdot \mathbf{r}}\times\nonumber
\\
&&\left\{ m_{X}(\mathbf{G})
\xi(\mathbf{G}+\mathbf{G}^{\prime},\mathbf{k})^2\frac{\eta(\mathbf{G}+\mathbf{G}^{\prime}+\mathbf{k})}{\left\vert  \mathbf{G}+\mathbf{G}^{\prime}+\mathbf{k}\right\vert^2}\right.\nonumber
\\
&+&\left.i m_{Y}(\mathbf{G})\xi(\mathbf{G}+\mathbf{G}^{\prime},\mathbf{k})
\frac{\eta(\mathbf{G}+\mathbf{G}^{\prime}+\mathbf{k})}{\left\vert  \mathbf{G}+\mathbf{G}^{\prime}+\mathbf{k}\right\vert}\right\}.
\label{hXX}
\end{eqnarray}

The coefficients $C_{\mathbf{G}}$ accounts the geometry of the periodic structure, which may be in the form of stripes, circular dots, squares, etc.\cite{Gallardo14}
In general, the static field component $H_{Z}^{I0}(\mathbf{r})$ and $h^{I}_{X,Y}$ enter directly in the dynamics of the system through Eq. (\ref{LL1}), while the $H_{X}^{I0}(\mathbf{r})$ and $H_{Y}^{I0}(\mathbf{r})$ components affect the static properties of the system, as will be discussed in Sec. \ref{SIII}.
 
%---------------FIGURE 2-------------
\begin{figure*}[t]
\includegraphics[width=1\textwidth]{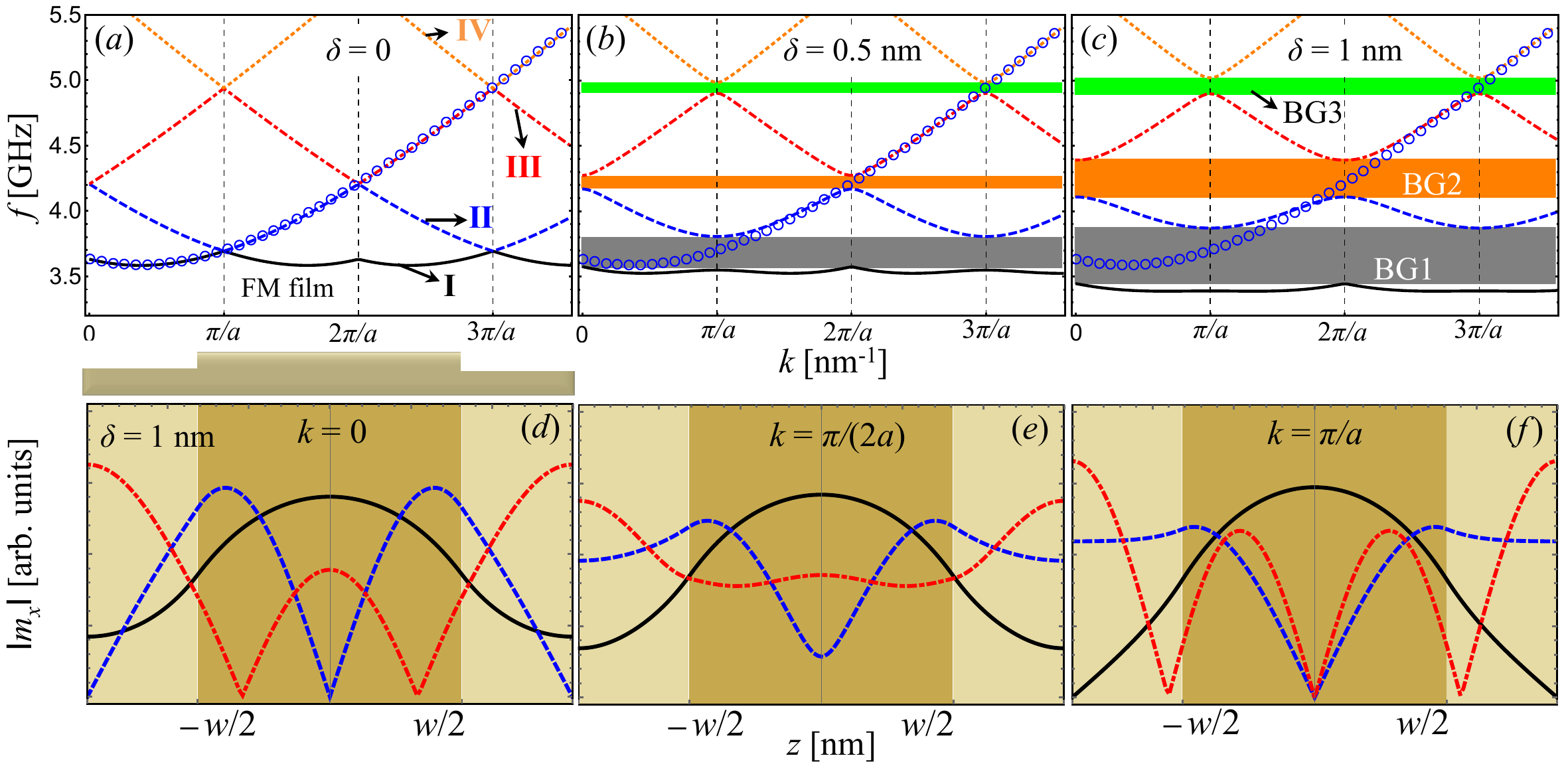} 
\caption{($a$), ($b$) and ($c$) show the dispersion relation in the backward volume configuration for $\mu_0H=$ 15 mT and $\delta=$ 0, 0.5 and 1 nm, respectively.
Also, the circles depict the dispersion of the thin film without periodic modulation. 
($d$), ($e$) and ($f$) depict spin-wave profiles at $\delta=1$ nm for the cases $k=0$, $k=\pi/(2a)$ and $k=\pi/a$, respectively.  
Note that the modes have been labeled as I for the lower frequency mode and II, III, etc. for the upper frequency ones.}
\label{Fig2}
\end{figure*}
%---------------FIGURE 2---------------
Now, inserting all field contributions into Eq. (\ref{LL1}), the following eigenvalue problem is obtained: 
\begin{eqnarray}
\tilde{\mathbf{A}}\ \mathbf{m}^{\rm T}_{\mathbf{G}}&=&i\Omega \ \mathbf{m}^{\rm T}_{\mathbf{G}}
\label{BT}
\end{eqnarray}
where $\mathbf{m}^{\rm T}_{\mathbf{G}}=[m_X(G_1),...,m_X(G_N),m_Y(G_1),...,m_Y(G_N)]$ is the eigenvector and $\tilde{\mathbf{A}}$ is given by
\begin{eqnarray}
\tilde{\mathbf{A}}=\left( 
\begin{array}{cc}
 \tilde{\mathbf{A}}^{XX} 
& \tilde{\mathbf{A}}^{XY} 
\\
 \tilde{\mathbf{A}}^{YX} 
& \tilde{\mathbf{A}}^{YY} 
\end{array}\right).
\label{subM}
\end{eqnarray}
After a calculation the submatrices in Eq.\ (\ref{subM}) are given by
\begin{widetext}
\begin{subequations}
\begin{eqnarray}
 \mathbf{A}^{XX}_{\mathbf{G},\mathbf{G}^{\prime}}  &=&
 - \mathbf{A}^{YY}_{\mathbf{G},\mathbf{G}^{\prime}}=-i 2\pi M_s C_{\mathbf{G}-\mathbf{G}^{\prime}} \xi(\mathbf{G},\mathbf{k})
\frac{\eta(\mathbf{G}+\mathbf{k})}{\left\vert  \mathbf{G}+\mathbf{k}\right\vert}
\\
 \mathbf{A}^{XY}_{\mathbf{G},\mathbf{G}^{\prime}}  &=&- \left[D_{ex}(\mathbf{G}+\mathbf{k})^2 +4\pi M_s\zeta(\mathbf{G})+H\cos\varphi\right]\delta_{\mathbf{G},\mathbf{G}^{\prime}}+F^{I}_{XY},
\\
 \mathbf{A}^{YX}_{\mathbf{G},\mathbf{G}^{\prime}}   &=& \left[D_{ex}(\mathbf{G}+\mathbf{k})^2 +4\pi M_s\xi(\mathbf{G},\mathbf{k})^2 \frac{1-\zeta(\mathbf{G})}{\vert \mathbf{G}+\mathbf{k}\vert^2}+H\cos\varphi\right]\delta_{\mathbf{G},\mathbf{G}^{\prime}}+F^{I}_{YX}.
\end{eqnarray}
\label{AA}
\end{subequations}
\end{widetext}
Here, the functions $F^{I}_{XY}$ and $F^{I}_{YX}$ come from the dipolar interaction between the etched zone and the thick part and are given by:
\begin{equation}
F^{I}_{XY}=2\pi M_{s} C_{\mathbf{G}-\mathbf{G}^{\prime}}\left[\chi(\mathbf{G}-\mathbf{G}^{\prime})^2 \eta(\mathbf{G}-\mathbf{G}^{\prime})+\eta(\mathbf{G}+\mathbf{k})\right],
\label{BT1}
\end{equation}
and
\begin{eqnarray}
F^{I}_{YX}&=&2\pi M_{s} C_{\mathbf{G}-\mathbf{G}^{\prime}}\left[-\chi(\mathbf{G}-\mathbf{G}^{\prime})^2 \eta(\mathbf{G}-\mathbf{G}^{\prime})\right.\nonumber
\\ &+& 
\left.\xi(\mathbf{G},\mathbf{k})^2\frac{\eta(\mathbf{G}+\mathbf{k})}{\left\vert  \mathbf{G}+\mathbf{k}\right\vert^2}\right].
\label{BT2}
\end{eqnarray}
Then, by using standard numerical methods and a convergence test to check the reliability of the results, the eigenvalues  and eigenvectors of Eq.\ (\ref{BT}) can be obtained.
%%%%%%%%%%%%%%THEORY%%%%%%%%%%%%%%
%---------------FIGURE NEW---------------
\begin{figure}[h]
\includegraphics[width=8.5 cm]{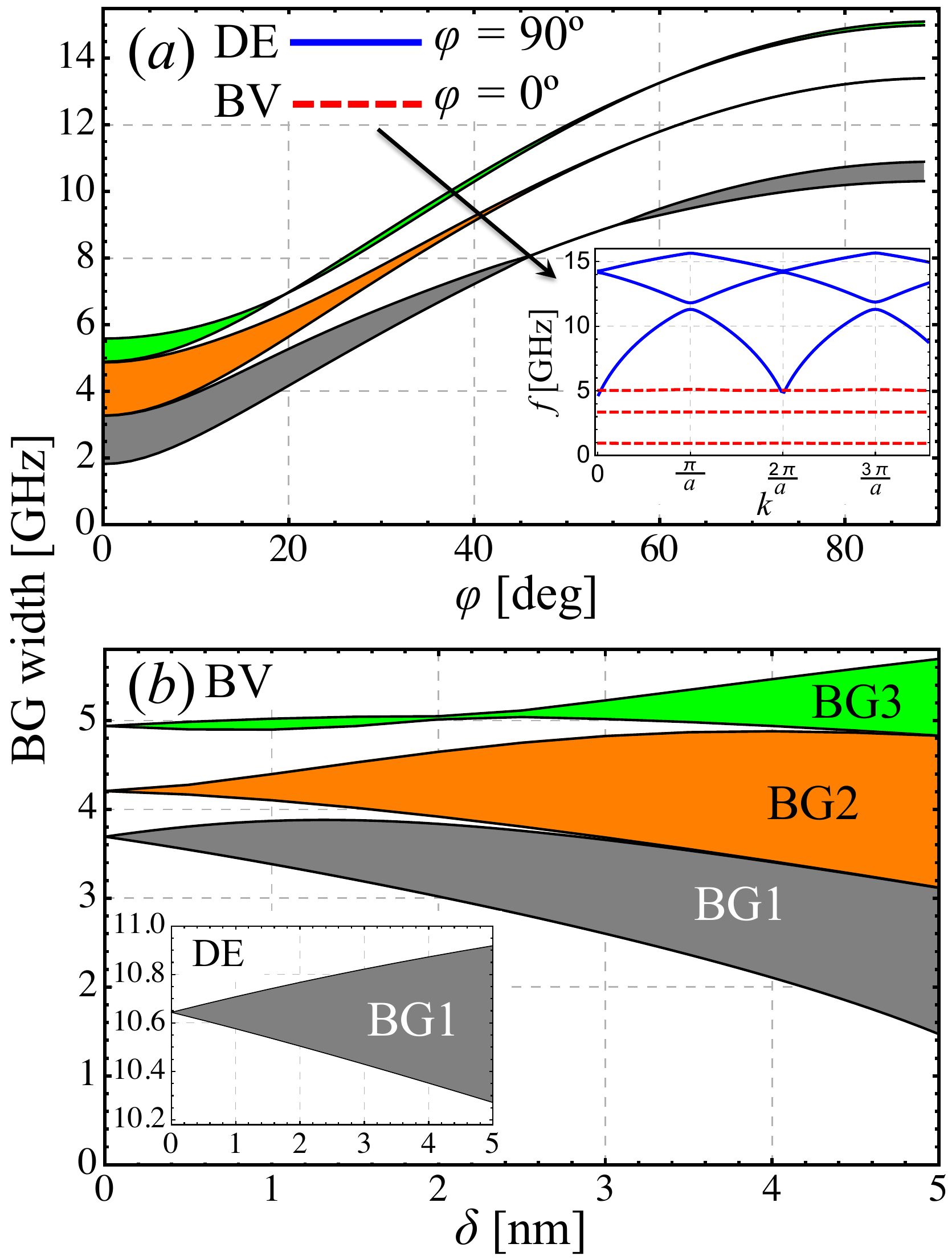}
\caption{In ($a$), the in-plane angular dependence of the first three band gaps in surface-modulated magnonic crystals with a larger defect depth, $\delta=4.5$ nm are shown.
Here, the inset shows the dispersion relation for the Damon-Eshbach modes (solid lines) and backward volume modes (dashed lines).
Fig. ($b$) depicts band gaps width as a function of the depth $\delta$ in the BV geometry, while the inset shows the first gap against $\delta$ for DE configuration.}
\label{FigNEW}
\end{figure}
%---------------FIGURE NEW---------------
%%%%%%%%%%%%%%RESULTS AND DISCUSSION%%%%%%%%%%%%%%
\section{Results and discussion}
\label{SIII}

The theoretical model will be applied now to thin films with one-dimensional stripe-like modulations, as shown in Fig. \ref{Fig1}.
For such geometry, the Fourier coefficient is given by $C_{\mathbf{G}}=(w/a){\rm{sinc}}[(w/a)\pi n]$, where $w$ and $a$ are previously defined in Fig. \ref{Fig1}($b$).
Also, at $N$ = 50, a convergence of the numerical solutions is reached.
Typical permalloy parameters are used, namely a saturation magnetization $M_{s}=797$ kA/m, stiffness constant $D_{ex}=24.67$ T nm$^{2}$ and the gyromagnetic ratio $\gamma =184.764$ GHz/T.
Moreover, the geometrical parameters of the etched zones are $d = 27$ nm, $a=299$ nm and $w=163$ nm, which are chosen for a comparison with micromagnetic simulations.

In Fig. \ref{Fig2}($a$)--($c$), dispersion relations in backward volume (BV) geometry ($\varphi=0$) for $\mu_0H=$15 mT and $\delta=$0, 0.5 and 1 nm are shown, respectively. 
The circles indicate the dispersion of the perfect film without periodic modulation.
Here, it is clearly visible that the periodic stray field $H^{0I}(\mathbf{r})$, created by the static magnetic charges, opens frequency band gaps, whose strength can be controlled through the depth $\delta$ of the surface modulation.
Note that the widths of the first three frequency band gaps are given by ${\rm{BG1}}=f^{(\rm{II})}(\pi/a)-f^{\rm{I}}(0)$, ${\rm{BG2}}=f^{(\rm{III})}(2\pi/a)-f^{\rm{II}}(2\pi/a)$, and ${\rm{BG3}}=f^{(\rm{IV})}(3\pi/a)-f^{\rm{III}}(3\pi/a)$, as shown Fig. \ref{Fig2}($b$)--($c$). 
The first band gap BG1 is an indirect gap, where the lower frequency mode [solid lines in Fig. \ref{Fig2}($b$)--($c$)]  keeps a finite group velocity at $k=0$.
The spatial profiles of spin waves for the case $\delta=1$ nm are shown in Fig. \ref{Fig2}($d$)--($f$), where the wave vectors $k=0$, $\pi/(2a)$ and $\pi/a$ were selected.
One can see that in the first Brillouin zone ($k=\pi/a$), the three lower modes 1, 2 and 3 are standing modes, namely the group velocity for all of them is zero, as shown Figs. \ref{Fig2}($c$) and \ref{Fig2}($f$). 
In Fig. \ref{Fig2}($e$), the group velocity is non zero and therefore spin-wave propagation is present, either with positive or negative group velocities.
At $k=0$, the lower mode presents a non-zero group velocity, while the group velocity of the two upper ones is zero.
Furthermore, one can see that in Fig. \ref{Fig2}($d$) the dashed mode has an antisymmetric behavior, while the dot-dashed and solid ones are symmetric, thus, the detection of either symmetric or antisymmetric modes or both depends on the excitation geometry. For conventional FMR and BLS measurements, most of the time, the detection is restricted on the symmetric modes only.

 According to the model presented, band gaps are opened by the static and dynamic dipole fields created respectively by the static and dynamic magnetizations. 
In the linear regime the gaps in backward volume configuration are induced only by static magnetic charges, while the gaps for Damon-Eshbach (DE) spin waves are originating from the dynamic magnetic charges. 
Interestingly, at intermediate angles between $\mathbf{M}$ and $\mathbf{k}$, both kinds of magnetic charges contribute to the opening of the magnonic band gaps. 
This is evident from Fig. \ref{FigNEW}($a$), where the widths of the first three band gaps are plotted as a function of the in-plane angle in a surface-modulated magnonic crystal with a depth $\delta=4.5$ nm. 
Here, one can clearly see that band gap widths induced by static magnetic charges (at $\varphi=0$) are substantially larger than the BG widths induced by the dynamic magnetic charges (at $\varphi=\pi/2$). 
Note that there are intermediate angles (around $\varphi \approx 50-70$ deg) where the BG width vanishes.
The inset in \ref{FigNEW}($a$) shows the dispersion relations of the SWs for both DE modes (solid lines) and BV modes (dashed lines). 
Note that the BV configuration shows almost dispersionless modes, since the frequency BGs are strong enough to suppress the SW propagation, confining the SW modes to some narrow frequency bands.
Dispersionless SW modes have been observed in magnetic microwire arrays in the DE configuration, where the geometrical confinement of the spin waves leads to the quantization of the modes.\cite{Mathieu98,Jorzick99}

The evolution of the BG width as a function of modulation depth $\delta$ is shown in Fig \ref{FigNEW}($b$) for SWs in the BV geometry. 
Overall, it is possible to see that the gaps increase with $\delta$, because the magnetic dipole fields that create the gaps become stronger for larger modulation depths.
Nevertheless, a peculiar modulation of the third band gap is noted, since the BG3 decreases up to $\delta=2$ nm and then increases again.
In DE configuration the BGs slightly increases almost linearly with $\delta$, as shown the inset in Fig. \ref{FigNEW}($b$), where the first BG is depicted.
It can be concluded that the BGs in backward volume configuration are considerable larger than the gaps for surface waves, which is clearly visible for any modulation depth. 

According to Eq. (\ref{AA}), it is possible to show that if the SW profile along the thickness is uniform (dynamic magnetization components are independent of $y$-axis), the frequency of spin waves only depends on the square of $\xi(\mathbf{G},\mathbf{k})$ and,  hence, two counterpropagating spin waves are present, which exhibit a full reciprocity, namely $f(\mathbf{k})=f(-\mathbf{k})$.
This can be demonstrated from the diagonal elements $\mathbf{A}^{XX}_{\mathbf{G},\mathbf{G}^{\prime}}$ and $\mathbf{A}^{YY}_{\mathbf{G},\mathbf{G}^{\prime}}$ defined in Eq. (\ref{AA}a), which are dependent on the sign of the wave vector through function $\xi(\mathbf{G})$ defined in Eq. (\ref{Xi}).
Nevertheless, if the so-called first perpendicular standing spin-wave mode, which has an antisymmetric profile across the film thickness, is taken into account the SW frequency becomes dependent of the wave vector orientation and non-reciprocal features appear, i.e. $f(\mathbf{k})\neq f(-\mathbf{k})$.
This effect has been observed in Refs. \onlinecite{Dib15} and \onlinecite{Gladii16} for FM films with different top and bottom surfaces. 
Note that in the one-dimensional case $\xi(\mathbf{G},\mathbf{k})=(G_n+k_z)\sin\varphi$ and therefore the non-reciprocal properties are enhanced in the Damon-Eshbach geometry ($\varphi=90^{\circ}$).  
Such non-reciprocal features of spin waves are important since they allow for performing logic operations that may be useful for insulators and circulators,\cite{Grundler15} and such non-reciprocity has recently been observed in thin films with Dzyaloshinskii--Moriya interaction.\cite{Zakeri10,Cortes13,Kai15,Zhang15,Cho15,Nembach15,Belmeguenai15}

\begin{figure}[h]
\includegraphics[width=8.5 cm]{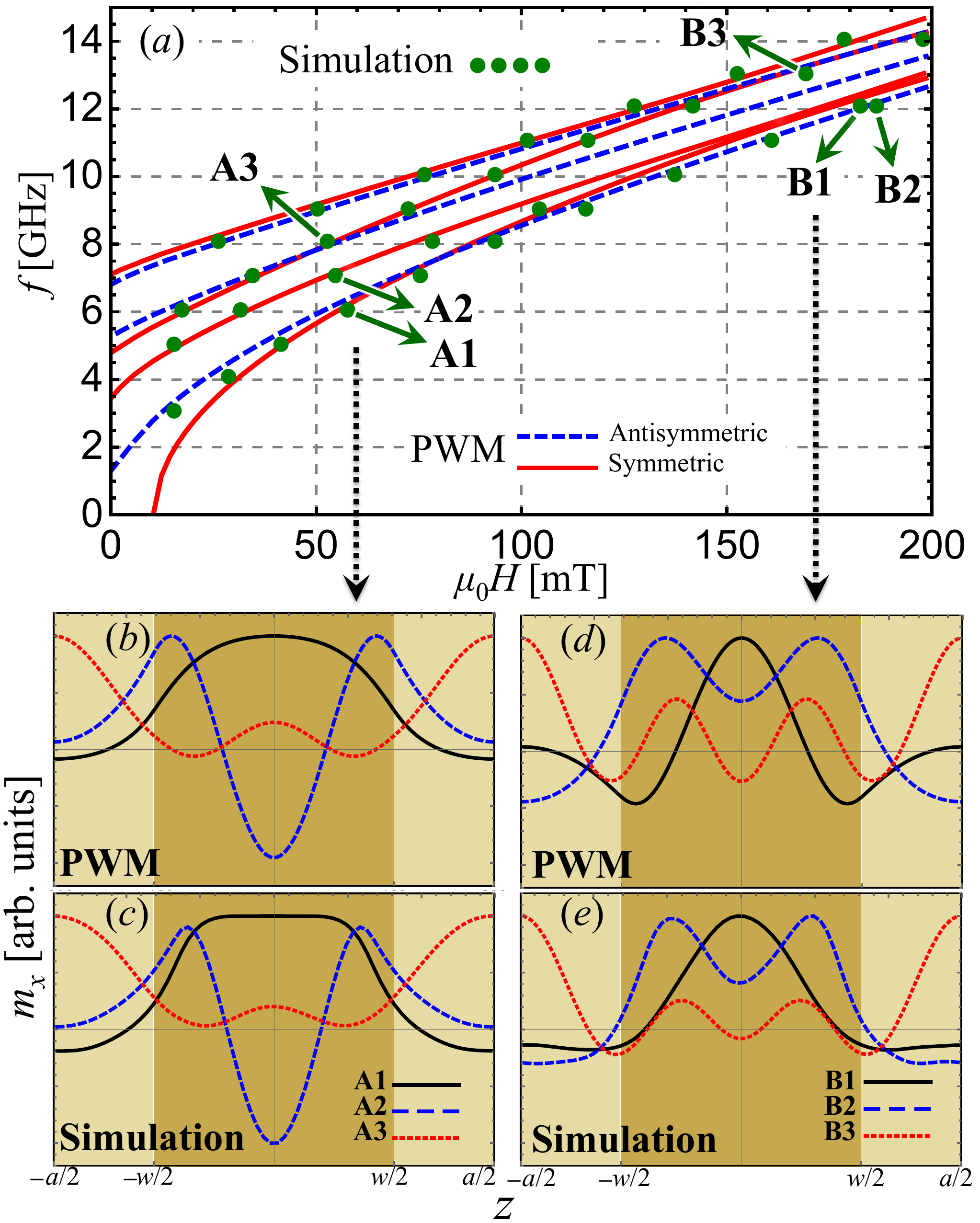}
\caption{Results from the plane wave-based theory and the FMR simulations are shown for $\delta=$ 4.5 nm.
In ($a$) the resonance frequency as a function of field is shown by using both methods.
The solid (dashed) lines depict the symmetric (antisymmetric) modes.
Figs. ($b$) and ($d$) depict theoretical results for the spin-wave amplitude $m_x$, while in ($c$) and ($e$) the numerical counterpart is shown.}
\label{Fig3}
\end{figure}

Additionally, to get insight about both the frequency-dependence of the modes and the SW profiles in the long wavelength limit, micromagnetic simulations have been carried out using the \emph{MuMax$^3$} code.\cite{Vansteenkiste14} 
Here, a magnetic film was built up in the ($x$;\;$y$;\;$z$) dimensions (100~nm;\;27~nm;\;299~nm) with a mesh size of (4.5~nm;\;4.672~nm;\;6.25~nm). Next, a 163~nm wide wire of 4.5~nm thickness was centrally put on top of the film forming the intact film part. To consider the reality of an extended surface modulated film, periodic boundary conditions were chosen along the $x$- and $z$-directions. The external field was applied in the $z$-direction whereas the excitation field was chosen in $x$-direction. The simulation of the FMR response was carried out according to the approach presented in Ref.~\onlinecite{Wagner15} with a continuous wave excitation for a swept external field at a given frequency. 
The dynamic magnetization component $m_x$ can be employed to obtain both, the FMR response $m_x(H)$ for a given frequency as well as the respective spin-wave profile $m_x(z)$. 
The magnetic parameters are the same used in the analytical approach, with a damping constant $\alpha = 6.5\times 10^{-3}$.
%The magnetic material parameters are $M_s = 797$~kA\,m$^{-1}$, $A = 9.83$~pJ\,m$^{-1}$, $\gamma/2\pi = 184.764$~GHz\,T$^{-1}$ and $\alpha = 6.5\times 10^{-3}$.
Fig. \ref{Fig3}($a$) shows a comparison between theory and numerical simulations, where the mode frequency is shown depending on the external field $\mu_0 H$, demonstrating the reliability of the developed theory based on the plane-wave method.\cite{Sokolovskyy11}
Such behavior was previously obtained using linear response theory and two-magnon scattering in the limit of perturbative modulation depths.\cite{Landeros12,Gallardo14} 
It is worth mentioning that in Fig. \ref{Fig3}($a$) no fitting parameters were used, since all geometrical and magnetic parameters have been used accordingly to the simulation input.
Noticeably, there is a good agreement between both approaches.
The mode-profiles extracted from numerical simulations can be directly compared to the ones obtained from the theoretical model as shown Figs. \ref{Fig3}($b$)--($e$).
In Figs. \ref{Fig3}($b$) and ($c$), the spin-wave amplitudes $m_x$  for the modes A1, A2 and A3 are shown, where both theory and simulations manifest a good agreement.
Notice that the three modes are symmetric, since the excitation of antisymmetric modes requires an inhomogeneous excitation.  
Here, modes A1 and A2 are mainly localized in the thicker part of the periodic structure, while the third one is localized in the thinner part.
This behavior can be explained by the periodic modulation of the stray field shown in Fig. \ref{Fig1}($c$), acting as demagnetizing (magnetizing) field for the local magnetization in the thicker (thinner) part.
Therefore, at a fixed applied field the internal field in the thick (thin) part decreases (increases) the effective field, such that the mode is shifted to lower (higher) frequencies. 

%---------------FIGURE 4---------------
%---------------FIGURE 4---------------

It is worth highlighting that the good agreement between the FMR numerical simulations and the results from the theoretical plane-wave method regarding the frequency-dependence as well as the spin-wave profiles allows for further interpretations. 
The theoretical model provides access to the $k \neq 0$ spin waves of the system and thus, for estimating the position and width of band gaps. 
The benefit would be to circumvent complicated BLS measurements, which is one of the main techniques that provides access to such kind of information.

%%%%%%%%%%%%%%RESULTS AND DISCUSSION%%%%%%%%%%%%%%

%%%%%%%%%%%CONCLUSIONS%%%%%%%%%%%%
\section{Conclusions}
\label{SIV}
Spin waves in surface-modulated ferromagnetic thin films were theoretically studied using a model based on the plane-wave method and micromagnetic simulations. 
The theory shows that the dipolar interaction produced by surface geometrical modulation is capable to open magnonic band gaps either in the backward volume or Damon-Eshbach configurations, whose magnitude can be controlled by the etching height.
Band-gap widths created by static magnetic charges is found to be broader than the one created by dynamic magnetic charges.
The approach agrees very well with FMR numerical simulations in the long wavelength limit, which allows to  validate the theoretical model.
A comparison with numerical simulations of the frequency versus field dependence as well as the spin-wave profiles was conducted showing a good agreement. 
Consequently, the model applied to periodically etched thin films provides further key-information about band gaps modulation, spatial localization of the modes and the dispersion of the spin waves.
Therefore, the results obtained in this work offer a better understanding of such systems paving the way for further developments of magnonic crystal based devices.
%%%%%%%%%CONCLUSIONS%%%%%%%%%%%%
\\

%%%%%%%%ACKNOWLEDGMENTS%%%%%%%%%
We acknowledge financial support from CONICYT PAI/ACADEMIA 79140033,  CONICYT/DAAD PCCI140051, FONDECYT 1161403, and Centers of excellence with Basal/CONICYT financing, grant FB0807, CEDENNA.
Funding from Deutsche Forschungsgemeinschaft (grant no. 2443/5-1) and DAAD PPP ALECHILE (grant no. 57136331) is highly acknowledged.

%%%%%%%%%%%%%%ACKNOWLEDGMENTS%%%%%%%%%%%%%%

\bibliographystyle{apsrev4-1}
\bibliography{bibliografiaEP2.0}

%merlin.mbs apsrev4-1.bst 2010-07-25 4.21a (PWD, AO, DPC) hacked
%Control: key (0)
%Control: author (72) initials jnrlst
%Control: editor formatted (1) identically to author
%Control: production of article title (-1) disabled
%Control: page (0) single
%Control: year (1) truncated
%Control: production of eprint (0) enabled
\begin{thebibliography}{63}%
\makeatletter
\providecommand \@ifxundefined [1]{%
 \@ifx{#1\undefined}
}%
\providecommand \@ifnum [1]{%
 \ifnum #1\expandafter \@firstoftwo
 \else \expandafter \@secondoftwo
 \fi
}%
\providecommand \@ifx [1]{%
 \ifx #1\expandafter \@firstoftwo
 \else \expandafter \@secondoftwo
 \fi
}%
\providecommand \natexlab [1]{#1}%
\providecommand \enquote  [1]{``#1''}%
\providecommand \bibnamefont  [1]{#1}%
\providecommand \bibfnamefont [1]{#1}%
\providecommand \citenamefont [1]{#1}%
\providecommand \href@noop [0]{\@secondoftwo}%
\providecommand \href [0]{\begingroup \@sanitize@url \@href}%
\providecommand \@href[1]{\@@startlink{#1}\@@href}%
\providecommand \@@href[1]{\endgroup#1\@@endlink}%
\providecommand \@sanitize@url [0]{\catcode `\\12\catcode `\$12\catcode
  `\&12\catcode `\#12\catcode `\^12\catcode `\_12\catcode `\%12\relax}%
\providecommand \@@startlink[1]{}%
\providecommand \@@endlink[0]{}%
\providecommand \url  [0]{\begingroup\@sanitize@url \@url }%
\providecommand \@url [1]{\endgroup\@href {#1}{\urlprefix }}%
\providecommand \urlprefix  [0]{URL }%
\providecommand \Eprint [0]{\href }%
\providecommand \doibase [0]{http://dx.doi.org/}%
\providecommand \selectlanguage [0]{\@gobble}%
\providecommand \bibinfo  [0]{\@secondoftwo}%
\providecommand \bibfield  [0]{\@secondoftwo}%
\providecommand \translation [1]{[#1]}%
\providecommand \BibitemOpen [0]{}%
\providecommand \bibitemStop [0]{}%
\providecommand \bibitemNoStop [0]{.\EOS\space}%
\providecommand \EOS [0]{\spacefactor3000\relax}%
\providecommand \BibitemShut  [1]{\csname bibitem#1\endcsname}%
\let\auto@bib@innerbib\@empty
%</preamble>
\bibitem [{\citenamefont {Demokritov}\ and\ \citenamefont
  {Slavin}(2012)}]{Demokritov12}%
  \BibitemOpen
  \bibfield  {author} {\bibinfo {author} {\bibfnamefont {S.}~\bibnamefont
  {Demokritov}}\ and\ \bibinfo {author} {\bibfnamefont {A.}~\bibnamefont
  {Slavin}},\ }\href {https://books.google.cl/books?id=CK-3S9Xa-z4C} {\emph
  {\bibinfo {title} {Magnonics: From Fundamentals to Applications}}},\ Topics
  in Applied Physics\ (\bibinfo  {publisher} {Springer Berlin Heidelberg},\
  \bibinfo {year} {2012})\BibitemShut {NoStop}%
\bibitem [{\citenamefont {Chumak}\ \emph {et~al.}(2015)\citenamefont {Chumak},
  \citenamefont {Vasyuchka}, \citenamefont {Serga},\ and\ \citenamefont
  {Hillebrands}}]{Chumak15}%
  \BibitemOpen
  \bibfield  {author} {\bibinfo {author} {\bibfnamefont {A.~V.}\ \bibnamefont
  {Chumak}}, \bibinfo {author} {\bibfnamefont {V.~I.}\ \bibnamefont
  {Vasyuchka}}, \bibinfo {author} {\bibfnamefont {A.~A.}\ \bibnamefont
  {Serga}}, \ and\ \bibinfo {author} {\bibfnamefont {B.}~\bibnamefont
  {Hillebrands}},\ }\href {http://dx.doi.org/10.1038/nphys3347} {\bibfield
  {journal} {\bibinfo  {journal} {Nat. Phys.}\ }\textbf {\bibinfo {volume}
  {11}},\ \bibinfo {pages} {453} (\bibinfo {year} {2015})}\BibitemShut
  {NoStop}%
\bibitem [{\citenamefont {Kajiwara}\ \emph {et~al.}(2010)\citenamefont
  {Kajiwara}, \citenamefont {Harii}, \citenamefont {Takahashi}, \citenamefont
  {Ohe}, \citenamefont {Uchida}, \citenamefont {Mizuguchi}, \citenamefont
  {Umezawa}, \citenamefont {Kawai}, \citenamefont {Ando}, \citenamefont
  {Takanashi}, \citenamefont {Maekawa},\ and\ \citenamefont
  {Saitoh}}]{Kajiwara10}%
  \BibitemOpen
  \bibfield  {author} {\bibinfo {author} {\bibfnamefont {Y.}~\bibnamefont
  {Kajiwara}}, \bibinfo {author} {\bibfnamefont {K.}~\bibnamefont {Harii}},
  \bibinfo {author} {\bibfnamefont {S.}~\bibnamefont {Takahashi}}, \bibinfo
  {author} {\bibfnamefont {J.}~\bibnamefont {Ohe}}, \bibinfo {author}
  {\bibfnamefont {K.}~\bibnamefont {Uchida}}, \bibinfo {author} {\bibfnamefont
  {M.}~\bibnamefont {Mizuguchi}}, \bibinfo {author} {\bibfnamefont
  {H.}~\bibnamefont {Umezawa}}, \bibinfo {author} {\bibfnamefont
  {H.}~\bibnamefont {Kawai}}, \bibinfo {author} {\bibfnamefont
  {K.}~\bibnamefont {Ando}}, \bibinfo {author} {\bibfnamefont {K.}~\bibnamefont
  {Takanashi}}, \bibinfo {author} {\bibfnamefont {S.}~\bibnamefont {Maekawa}},
  \ and\ \bibinfo {author} {\bibfnamefont {E.}~\bibnamefont {Saitoh}},\ }\href
  {http://dx.doi.org/10.1038/nature08876} {\bibfield  {journal} {\bibinfo
  {journal} {Nature}\ }\textbf {\bibinfo {volume} {464}},\ \bibinfo {pages}
  {262} (\bibinfo {year} {2010})}\BibitemShut {NoStop}%
\bibitem [{\citenamefont {Khitun}\ and\ \citenamefont {Wang}(2005)}]{Khitun05}%
  \BibitemOpen
  \bibfield  {author} {\bibinfo {author} {\bibfnamefont {A.}~\bibnamefont
  {Khitun}}\ and\ \bibinfo {author} {\bibfnamefont {K.~L.}\ \bibnamefont
  {Wang}},\ }\href {\doibase http://dx.doi.org/10.1016/j.spmi.2005.07.001}
  {\bibfield  {journal} {\bibinfo  {journal} {Superlattice. Microst.}\ }\textbf
  {\bibinfo {volume} {38}},\ \bibinfo {pages} {184 } (\bibinfo {year}
  {2005})}\BibitemShut {NoStop}%
\bibitem [{\citenamefont {Khitun}\ \emph {et~al.}(2010)\citenamefont {Khitun},
  \citenamefont {Bao},\ and\ \citenamefont {Wang}}]{Khitun10}%
  \BibitemOpen
  \bibfield  {author} {\bibinfo {author} {\bibfnamefont {A.}~\bibnamefont
  {Khitun}}, \bibinfo {author} {\bibfnamefont {M.}~\bibnamefont {Bao}}, \ and\
  \bibinfo {author} {\bibfnamefont {K.~L.}\ \bibnamefont {Wang}},\ }\href
  {http://stacks.iop.org/0022-3727/43/i=26/a=264005} {\bibfield  {journal}
  {\bibinfo  {journal} {J. Phys. D: Appl. Phys.}\ }\textbf {\bibinfo {volume}
  {43}},\ \bibinfo {pages} {264005} (\bibinfo {year} {2010})}\BibitemShut
  {NoStop}%
\bibitem [{\citenamefont {Jamali}\ \emph {et~al.}(2013)\citenamefont {Jamali},
  \citenamefont {Kwon}, \citenamefont {Seo}, \citenamefont {Lee},\ and\
  \citenamefont {Yang}}]{Jamali13}%
  \BibitemOpen
  \bibfield  {author} {\bibinfo {author} {\bibfnamefont {M.}~\bibnamefont
  {Jamali}}, \bibinfo {author} {\bibfnamefont {J.~H.}\ \bibnamefont {Kwon}},
  \bibinfo {author} {\bibfnamefont {S.-M.}\ \bibnamefont {Seo}}, \bibinfo
  {author} {\bibfnamefont {K.-J.}\ \bibnamefont {Lee}}, \ and\ \bibinfo
  {author} {\bibfnamefont {H.}~\bibnamefont {Yang}},\ }\href
  {http://dx.doi.org/10.1038/srep03160} {\bibfield  {journal} {\bibinfo
  {journal} {Sci. Rep.}\ }\textbf {\bibinfo {volume} {3}} (\bibinfo {year}
  {2013})}\BibitemShut {NoStop}%
\bibitem [{\citenamefont {Kostylev}\ \emph {et~al.}(2005)\citenamefont
  {Kostylev}, \citenamefont {Serga}, \citenamefont {Schneider}, \citenamefont
  {Leven},\ and\ \citenamefont {Hillebrands}}]{Kostylev05}%
  \BibitemOpen
  \bibfield  {author} {\bibinfo {author} {\bibfnamefont {M.~P.}\ \bibnamefont
  {Kostylev}}, \bibinfo {author} {\bibfnamefont {A.~A.}\ \bibnamefont {Serga}},
  \bibinfo {author} {\bibfnamefont {T.}~\bibnamefont {Schneider}}, \bibinfo
  {author} {\bibfnamefont {B.}~\bibnamefont {Leven}}, \ and\ \bibinfo {author}
  {\bibfnamefont {B.}~\bibnamefont {Hillebrands}},\ }\href
  {http://scitation.aip.org/content/aip/journal/apl/87/15/10.1063/1.2089147}
  {\bibfield  {journal} {\bibinfo  {journal} {Appl. Phys. Lett.}\ }\textbf
  {\bibinfo {volume} {87}},\ \bibinfo {eid} {153501} (\bibinfo {year}
  {2005})}\BibitemShut {NoStop}%
\bibitem [{\citenamefont {Tacchi}\ \emph {et~al.}(2015)\citenamefont {Tacchi},
  \citenamefont {Gruszecki}, \citenamefont {Madami}, \citenamefont {Carlotti},
  \citenamefont {K{\l}os}, \citenamefont {Krawczyk}, \citenamefont {Adeyeye},\
  and\ \citenamefont {Gubbiotti}}]{Tacchi15}%
  \BibitemOpen
  \bibfield  {author} {\bibinfo {author} {\bibfnamefont {S.}~\bibnamefont
  {Tacchi}}, \bibinfo {author} {\bibfnamefont {P.}~\bibnamefont {Gruszecki}},
  \bibinfo {author} {\bibfnamefont {M.}~\bibnamefont {Madami}}, \bibinfo
  {author} {\bibfnamefont {G.}~\bibnamefont {Carlotti}}, \bibinfo {author}
  {\bibfnamefont {J.~W.}\ \bibnamefont {K{\l}os}}, \bibinfo {author}
  {\bibfnamefont {M.}~\bibnamefont {Krawczyk}}, \bibinfo {author}
  {\bibfnamefont {A.}~\bibnamefont {Adeyeye}}, \ and\ \bibinfo {author}
  {\bibfnamefont {G.}~\bibnamefont {Gubbiotti}},\ }\href
  {http://dx.doi.org/10.1038/srep10367} {\bibfield  {journal} {\bibinfo
  {journal} {Sci. Rep.}\ }\textbf {\bibinfo {volume} {5}} (\bibinfo {year}
  {2015})}\BibitemShut {NoStop}%
\bibitem [{\citenamefont {Chumak}\ \emph {et~al.}(2014)\citenamefont {Chumak},
  \citenamefont {Serga},\ and\ \citenamefont {Hillebrands}}]{Chumak14}%
  \BibitemOpen
  \bibfield  {author} {\bibinfo {author} {\bibfnamefont {A.~V.}\ \bibnamefont
  {Chumak}}, \bibinfo {author} {\bibfnamefont {A.~A.}\ \bibnamefont {Serga}}, \
  and\ \bibinfo {author} {\bibfnamefont {B.}~\bibnamefont {Hillebrands}},\
  }\href {http://dx.doi.org/10.1038/ncomms5700} {\bibfield  {journal} {\bibinfo
   {journal} {Nat. Commun.}\ }\textbf {\bibinfo {volume} {5}},\ \bibinfo
  {pages} {4700} (\bibinfo {year} {2014})}\BibitemShut {NoStop}%
\bibitem [{\citenamefont {Vasseur}\ \emph {et~al.}(1996)\citenamefont
  {Vasseur}, \citenamefont {Dobrzynski}, \citenamefont {Djafari-Rouhani},\ and\
  \citenamefont {Puszkarski}}]{Vasseur96}%
  \BibitemOpen
  \bibfield  {author} {\bibinfo {author} {\bibfnamefont {J.~O.}\ \bibnamefont
  {Vasseur}}, \bibinfo {author} {\bibfnamefont {L.}~\bibnamefont {Dobrzynski}},
  \bibinfo {author} {\bibfnamefont {B.}~\bibnamefont {Djafari-Rouhani}}, \ and\
  \bibinfo {author} {\bibfnamefont {H.}~\bibnamefont {Puszkarski}},\ }\href
  {\doibase 10.1103/PhysRevB.54.1043} {\bibfield  {journal} {\bibinfo
  {journal} {Phys. Rev. B}\ }\textbf {\bibinfo {volume} {54}},\ \bibinfo
  {pages} {1043} (\bibinfo {year} {1996})}\BibitemShut {NoStop}%
\bibitem [{\citenamefont {Nikitov}\ \emph {et~al.}(2001)\citenamefont
  {Nikitov}, \citenamefont {Tailhades},\ and\ \citenamefont
  {Tsai}}]{Nikitov01}%
  \BibitemOpen
  \bibfield  {author} {\bibinfo {author} {\bibfnamefont {S.}~\bibnamefont
  {Nikitov}}, \bibinfo {author} {\bibfnamefont {P.}~\bibnamefont {Tailhades}},
  \ and\ \bibinfo {author} {\bibfnamefont {C.}~\bibnamefont {Tsai}},\ }\href
  {\doibase http://dx.doi.org/10.1016/S0304-8853(01)00470-X} {\bibfield
  {journal} {\bibinfo  {journal} {J. Magn. Magn. Mater.}\ }\textbf {\bibinfo
  {volume} {236}},\ \bibinfo {pages} {320 } (\bibinfo {year}
  {2001})}\BibitemShut {NoStop}%
\bibitem [{\citenamefont {H.~Puszkarski}(2003)}]{Puszkarski03}%
  \BibitemOpen
  \bibfield  {author} {\bibinfo {author} {\bibfnamefont {M.~K.}\ \bibnamefont
  {H.~Puszkarski}},\ }\href {\doibase 10.4028/www.scientific.net/SSP.94.125}
  {\bibfield  {journal} {\bibinfo  {journal} {Solid State Phenom.}\ }\textbf
  {\bibinfo {volume} {94}},\ \bibinfo {pages} {125} (\bibinfo {year}
  {2003})}\BibitemShut {NoStop}%
\bibitem [{\citenamefont {Kruglyak}\ and\ \citenamefont
  {Hicken}(2006)}]{Kruglyak06}%
  \BibitemOpen
  \bibfield  {author} {\bibinfo {author} {\bibfnamefont {V.}~\bibnamefont
  {Kruglyak}}\ and\ \bibinfo {author} {\bibfnamefont {R.}~\bibnamefont
  {Hicken}},\ }\href
  {http://www.sciencedirect.com/science/article/pii/S0304885306005889}
  {\bibfield  {journal} {\bibinfo  {journal} {J. Magn. Magn. Mater.}\ }\textbf
  {\bibinfo {volume} {306}},\ \bibinfo {pages} {191 } (\bibinfo {year}
  {2006})}\BibitemShut {NoStop}%
\bibitem [{\citenamefont {Gubbiotti}\ \emph {et~al.}(2007)\citenamefont
  {Gubbiotti}, \citenamefont {Tacchi}, \citenamefont {Carlotti}, \citenamefont
  {Singh}, \citenamefont {Goolaup}, \citenamefont {Adeyeye},\ and\
  \citenamefont {Kostylev}}]{Gubbiotti07}%
  \BibitemOpen
  \bibfield  {author} {\bibinfo {author} {\bibfnamefont {G.}~\bibnamefont
  {Gubbiotti}}, \bibinfo {author} {\bibfnamefont {S.}~\bibnamefont {Tacchi}},
  \bibinfo {author} {\bibfnamefont {G.}~\bibnamefont {Carlotti}}, \bibinfo
  {author} {\bibfnamefont {N.}~\bibnamefont {Singh}}, \bibinfo {author}
  {\bibfnamefont {S.}~\bibnamefont {Goolaup}}, \bibinfo {author} {\bibfnamefont
  {A.~O.}\ \bibnamefont {Adeyeye}}, \ and\ \bibinfo {author} {\bibfnamefont
  {M.}~\bibnamefont {Kostylev}},\ }\href
  {http://scitation.aip.org/content/aip/journal/apl/90/9/10.1063/1.2709909}
  {\bibfield  {journal} {\bibinfo  {journal} {Appl. Phys. Lett.}\ }\textbf
  {\bibinfo {volume} {90}},\ \bibinfo {eid} {092503} (\bibinfo {year}
  {2007})}\BibitemShut {NoStop}%
\bibitem [{\citenamefont {Krawczyk}\ and\ \citenamefont
  {Puszkarski}(2008)}]{Krawczyk08}%
  \BibitemOpen
  \bibfield  {author} {\bibinfo {author} {\bibfnamefont {M.}~\bibnamefont
  {Krawczyk}}\ and\ \bibinfo {author} {\bibfnamefont {H.}~\bibnamefont
  {Puszkarski}},\ }\href {\doibase 10.1103/PhysRevB.77.054437} {\bibfield
  {journal} {\bibinfo  {journal} {Phys. Rev. B}\ }\textbf {\bibinfo {volume}
  {77}},\ \bibinfo {pages} {054437} (\bibinfo {year} {2008})}\BibitemShut
  {NoStop}%
\bibitem [{\citenamefont {Neusser}\ \emph {et~al.}(2008)\citenamefont
  {Neusser}, \citenamefont {Botters},\ and\ \citenamefont
  {Grundler}}]{Neusser08}%
  \BibitemOpen
  \bibfield  {author} {\bibinfo {author} {\bibfnamefont {S.}~\bibnamefont
  {Neusser}}, \bibinfo {author} {\bibfnamefont {B.}~\bibnamefont {Botters}}, \
  and\ \bibinfo {author} {\bibfnamefont {D.}~\bibnamefont {Grundler}},\ }\href
  {\doibase 10.1103/PhysRevB.78.054406} {\bibfield  {journal} {\bibinfo
  {journal} {Phys. Rev. B}\ }\textbf {\bibinfo {volume} {78}},\ \bibinfo
  {pages} {054406} (\bibinfo {year} {2008})}\BibitemShut {NoStop}%
\bibitem [{\citenamefont {Chumak}\ \emph
  {et~al.}(2009{\natexlab{a}})\citenamefont {Chumak}, \citenamefont {Pirro},
  \citenamefont {Serga}, \citenamefont {Kostylev}, \citenamefont {Stamps},
  \citenamefont {Schultheiss}, \citenamefont {Vogt}, \citenamefont
  {Hermsdoerfer}, \citenamefont {Laegel}, \citenamefont {Beck},\ and\
  \citenamefont {Hillebrands}}]{Chumak09}%
  \BibitemOpen
  \bibfield  {author} {\bibinfo {author} {\bibfnamefont {A.~V.}\ \bibnamefont
  {Chumak}}, \bibinfo {author} {\bibfnamefont {P.}~\bibnamefont {Pirro}},
  \bibinfo {author} {\bibfnamefont {A.~A.}\ \bibnamefont {Serga}}, \bibinfo
  {author} {\bibfnamefont {M.~P.}\ \bibnamefont {Kostylev}}, \bibinfo {author}
  {\bibfnamefont {R.~L.}\ \bibnamefont {Stamps}}, \bibinfo {author}
  {\bibfnamefont {H.}~\bibnamefont {Schultheiss}}, \bibinfo {author}
  {\bibfnamefont {K.}~\bibnamefont {Vogt}}, \bibinfo {author} {\bibfnamefont
  {S.~J.}\ \bibnamefont {Hermsdoerfer}}, \bibinfo {author} {\bibfnamefont
  {B.}~\bibnamefont {Laegel}}, \bibinfo {author} {\bibfnamefont {P.~A.}\
  \bibnamefont {Beck}}, \ and\ \bibinfo {author} {\bibfnamefont
  {B.}~\bibnamefont {Hillebrands}},\ }\href
  {http://scitation.aip.org/content/aip/journal/apl/95/26/10.1063/1.3279138}
  {\bibfield  {journal} {\bibinfo  {journal} {Appl. Phys. Lett.}\ }\textbf
  {\bibinfo {volume} {95}},\ \bibinfo {eid} {262508} (\bibinfo {year}
  {2009}{\natexlab{a}})}\BibitemShut {NoStop}%
\bibitem [{\citenamefont {Lee}\ \emph {et~al.}(2009)\citenamefont {Lee},
  \citenamefont {Han},\ and\ \citenamefont {Kim}}]{Lee09}%
  \BibitemOpen
  \bibfield  {author} {\bibinfo {author} {\bibfnamefont {K.-S.}\ \bibnamefont
  {Lee}}, \bibinfo {author} {\bibfnamefont {D.-S.}\ \bibnamefont {Han}}, \ and\
  \bibinfo {author} {\bibfnamefont {S.-K.}\ \bibnamefont {Kim}},\ }\href
  {\doibase 10.1103/PhysRevLett.102.127202} {\bibfield  {journal} {\bibinfo
  {journal} {Phys. Rev. Lett.}\ }\textbf {\bibinfo {volume} {102}},\ \bibinfo
  {pages} {127202} (\bibinfo {year} {2009})}\BibitemShut {NoStop}%
\bibitem [{\citenamefont {Neusser}\ and\ \citenamefont
  {Grundler}(2009)}]{Neusser09}%
  \BibitemOpen
  \bibfield  {author} {\bibinfo {author} {\bibfnamefont {S.}~\bibnamefont
  {Neusser}}\ and\ \bibinfo {author} {\bibfnamefont {D.}~\bibnamefont
  {Grundler}},\ }\href {\doibase 10.1002/adma.200900809} {\bibfield  {journal}
  {\bibinfo  {journal} {Adv. Mater.}\ }\textbf {\bibinfo {volume} {21}},\
  \bibinfo {pages} {2927} (\bibinfo {year} {2009})}\BibitemShut {NoStop}%
\bibitem [{\citenamefont {Wang}\ \emph {et~al.}(2009)\citenamefont {Wang},
  \citenamefont {Zhang}, \citenamefont {Lim}, \citenamefont {Ng}, \citenamefont
  {Kuok}, \citenamefont {Jain},\ and\ \citenamefont {Adeyeye}}]{Wang09}%
  \BibitemOpen
  \bibfield  {author} {\bibinfo {author} {\bibfnamefont {Z.~K.}\ \bibnamefont
  {Wang}}, \bibinfo {author} {\bibfnamefont {V.~L.}\ \bibnamefont {Zhang}},
  \bibinfo {author} {\bibfnamefont {H.~S.}\ \bibnamefont {Lim}}, \bibinfo
  {author} {\bibfnamefont {S.~C.}\ \bibnamefont {Ng}}, \bibinfo {author}
  {\bibfnamefont {M.~H.}\ \bibnamefont {Kuok}}, \bibinfo {author}
  {\bibfnamefont {S.}~\bibnamefont {Jain}}, \ and\ \bibinfo {author}
  {\bibfnamefont {A.~O.}\ \bibnamefont {Adeyeye}},\ }\href
  {http://scitation.aip.org/content/aip/journal/apl/94/8/10.1063/1.3089839}
  {\bibfield  {journal} {\bibinfo  {journal} {Appl. Phys. Lett.}\ }\textbf
  {\bibinfo {volume} {94}},\ \bibinfo {eid} {083112} (\bibinfo {year}
  {2009})}\BibitemShut {NoStop}%
\bibitem [{\citenamefont {Kim}\ \emph {et~al.}(2009)\citenamefont {Kim},
  \citenamefont {Lee},\ and\ \citenamefont {Han}}]{Kim09}%
  \BibitemOpen
  \bibfield  {author} {\bibinfo {author} {\bibfnamefont {S.-K.}\ \bibnamefont
  {Kim}}, \bibinfo {author} {\bibfnamefont {K.-S.}\ \bibnamefont {Lee}}, \ and\
  \bibinfo {author} {\bibfnamefont {D.-S.}\ \bibnamefont {Han}},\ }\href
  {http://scitation.aip.org/content/aip/journal/apl/95/8/10.1063/1.3186782}
  {\bibfield  {journal} {\bibinfo  {journal} {Appl. Phys. Lett.}\ }\textbf
  {\bibinfo {volume} {95}},\ \bibinfo {eid} {082507} (\bibinfo {year}
  {2009})}\BibitemShut {NoStop}%
\bibitem [{\citenamefont {Serga}\ \emph {et~al.}(2010)\citenamefont {Serga},
  \citenamefont {Chumak},\ and\ \citenamefont {Hillebrands}}]{Serga10}%
  \BibitemOpen
  \bibfield  {author} {\bibinfo {author} {\bibfnamefont {A.~A.}\ \bibnamefont
  {Serga}}, \bibinfo {author} {\bibfnamefont {A.~V.}\ \bibnamefont {Chumak}}, \
  and\ \bibinfo {author} {\bibfnamefont {B.}~\bibnamefont {Hillebrands}},\
  }\href {http://stacks.iop.org/0022-3727/43/i=26/a=264002} {\bibfield
  {journal} {\bibinfo  {journal} {J. Phys. D: Appl. Phys.}\ }\textbf {\bibinfo
  {volume} {43}},\ \bibinfo {pages} {264002} (\bibinfo {year}
  {2010})}\BibitemShut {NoStop}%
\bibitem [{\citenamefont {Kruglyak}\ \emph {et~al.}(2010)\citenamefont
  {Kruglyak}, \citenamefont {Demokritov},\ and\ \citenamefont
  {Grundler}}]{Kruglyak10}%
  \BibitemOpen
  \bibfield  {author} {\bibinfo {author} {\bibfnamefont {V.~V.}\ \bibnamefont
  {Kruglyak}}, \bibinfo {author} {\bibfnamefont {S.~O.}\ \bibnamefont
  {Demokritov}}, \ and\ \bibinfo {author} {\bibfnamefont {D.}~\bibnamefont
  {Grundler}},\ }\href {http://stacks.iop.org/0022-3727/43/i=26/a=264001}
  {\bibfield  {journal} {\bibinfo  {journal} {J. Phys. D: Appl. Phys.}\
  }\textbf {\bibinfo {volume} {43}},\ \bibinfo {pages} {264001} (\bibinfo
  {year} {2010})}\BibitemShut {NoStop}%
\bibitem [{\citenamefont {Wang}\ \emph {et~al.}(2010)\citenamefont {Wang},
  \citenamefont {Zhang}, \citenamefont {Lim}, \citenamefont {Ng}, \citenamefont
  {Kuok}, \citenamefont {Jain},\ and\ \citenamefont {Adeyeye}}]{Wang10}%
  \BibitemOpen
  \bibfield  {author} {\bibinfo {author} {\bibfnamefont {Z.~K.}\ \bibnamefont
  {Wang}}, \bibinfo {author} {\bibfnamefont {V.~L.}\ \bibnamefont {Zhang}},
  \bibinfo {author} {\bibfnamefont {H.~S.}\ \bibnamefont {Lim}}, \bibinfo
  {author} {\bibfnamefont {S.~C.}\ \bibnamefont {Ng}}, \bibinfo {author}
  {\bibfnamefont {M.~H.}\ \bibnamefont {Kuok}}, \bibinfo {author}
  {\bibfnamefont {S.}~\bibnamefont {Jain}}, \ and\ \bibinfo {author}
  {\bibfnamefont {A.~O.}\ \bibnamefont {Adeyeye}},\ }\href
  {http://dx.doi.org/10.1021/nn901171u} {\bibfield  {journal} {\bibinfo
  {journal} {ACS Nano}\ }\textbf {\bibinfo {volume} {4}},\ \bibinfo {pages}
  {643} (\bibinfo {year} {2010})}\BibitemShut {NoStop}%
\bibitem [{\citenamefont {Gubbiotti}\ \emph {et~al.}(2010)\citenamefont
  {Gubbiotti}, \citenamefont {Tacchi}, \citenamefont {Madami}, \citenamefont
  {Carlotti}, \citenamefont {Adeyeye},\ and\ \citenamefont
  {Kostylev}}]{Gubbiotti10}%
  \BibitemOpen
  \bibfield  {author} {\bibinfo {author} {\bibfnamefont {G.}~\bibnamefont
  {Gubbiotti}}, \bibinfo {author} {\bibfnamefont {S.}~\bibnamefont {Tacchi}},
  \bibinfo {author} {\bibfnamefont {M.}~\bibnamefont {Madami}}, \bibinfo
  {author} {\bibfnamefont {G.}~\bibnamefont {Carlotti}}, \bibinfo {author}
  {\bibfnamefont {A.~O.}\ \bibnamefont {Adeyeye}}, \ and\ \bibinfo {author}
  {\bibfnamefont {M.}~\bibnamefont {Kostylev}},\ }\href
  {http://stacks.iop.org/0022-3727/43/i=26/a=264003} {\bibfield  {journal}
  {\bibinfo  {journal} {J. Phys. D: Appl. Phys.}\ }\textbf {\bibinfo {volume}
  {43}},\ \bibinfo {pages} {264003} (\bibinfo {year} {2010})}\BibitemShut
  {NoStop}%
\bibitem [{\citenamefont {Cao}\ \emph {et~al.}(2010)\citenamefont {Cao},
  \citenamefont {Yun}, \citenamefont {Liang},\ and\ \citenamefont
  {Bai}}]{Cao10}%
  \BibitemOpen
  \bibfield  {author} {\bibinfo {author} {\bibfnamefont {Y.}~\bibnamefont
  {Cao}}, \bibinfo {author} {\bibfnamefont {G.}~\bibnamefont {Yun}}, \bibinfo
  {author} {\bibfnamefont {X.}~\bibnamefont {Liang}}, \ and\ \bibinfo {author}
  {\bibfnamefont {N.}~\bibnamefont {Bai}},\ }\href
  {http://stacks.iop.org/0022-3727/43/i=30/a=305005} {\bibfield  {journal}
  {\bibinfo  {journal} {J. Phys. D: Appl. Phys.}\ }\textbf {\bibinfo {volume}
  {43}},\ \bibinfo {pages} {305005} (\bibinfo {year} {2010})}\BibitemShut
  {NoStop}%
\bibitem [{\citenamefont {Ding}\ \emph {et~al.}(2011)\citenamefont {Ding},
  \citenamefont {Kostylev},\ and\ \citenamefont {Adeyeye}}]{Ding11}%
  \BibitemOpen
  \bibfield  {author} {\bibinfo {author} {\bibfnamefont {J.}~\bibnamefont
  {Ding}}, \bibinfo {author} {\bibfnamefont {M.}~\bibnamefont {Kostylev}}, \
  and\ \bibinfo {author} {\bibfnamefont {A.~O.}\ \bibnamefont {Adeyeye}},\
  }\href {\doibase 10.1103/PhysRevB.84.054425} {\bibfield  {journal} {\bibinfo
  {journal} {Phys. Rev. B}\ }\textbf {\bibinfo {volume} {84}},\ \bibinfo
  {pages} {054425} (\bibinfo {year} {2011})}\BibitemShut {NoStop}%
\bibitem [{\citenamefont {Lenk}\ \emph {et~al.}(2011)\citenamefont {Lenk},
  \citenamefont {Ulrichs}, \citenamefont {Garbs},\ and\ \citenamefont
  {M{\"u}nzenberg}}]{Lenk11}%
  \BibitemOpen
  \bibfield  {author} {\bibinfo {author} {\bibfnamefont {B.}~\bibnamefont
  {Lenk}}, \bibinfo {author} {\bibfnamefont {H.}~\bibnamefont {Ulrichs}},
  \bibinfo {author} {\bibfnamefont {F.}~\bibnamefont {Garbs}}, \ and\ \bibinfo
  {author} {\bibfnamefont {M.}~\bibnamefont {M{\"u}nzenberg}},\ }\href
  {\doibase http://dx.doi.org/10.1016/j.physrep.2011.06.003} {\bibfield
  {journal} {\bibinfo  {journal} {Phys. Rep.}\ }\textbf {\bibinfo {volume}
  {507}},\ \bibinfo {pages} {107 } (\bibinfo {year} {2011})}\BibitemShut
  {NoStop}%
\bibitem [{\citenamefont {Tacchi}\ \emph {et~al.}(2011)\citenamefont {Tacchi},
  \citenamefont {Montoncello}, \citenamefont {Madami}, \citenamefont
  {Gubbiotti}, \citenamefont {Carlotti}, \citenamefont {Giovannini},
  \citenamefont {Zivieri}, \citenamefont {Nizzoli}, \citenamefont {Jain},
  \citenamefont {Adeyeye},\ and\ \citenamefont {Singh}}]{Tacchi11}%
  \BibitemOpen
  \bibfield  {author} {\bibinfo {author} {\bibfnamefont {S.}~\bibnamefont
  {Tacchi}}, \bibinfo {author} {\bibfnamefont {F.}~\bibnamefont {Montoncello}},
  \bibinfo {author} {\bibfnamefont {M.}~\bibnamefont {Madami}}, \bibinfo
  {author} {\bibfnamefont {G.}~\bibnamefont {Gubbiotti}}, \bibinfo {author}
  {\bibfnamefont {G.}~\bibnamefont {Carlotti}}, \bibinfo {author}
  {\bibfnamefont {L.}~\bibnamefont {Giovannini}}, \bibinfo {author}
  {\bibfnamefont {R.}~\bibnamefont {Zivieri}}, \bibinfo {author} {\bibfnamefont
  {F.}~\bibnamefont {Nizzoli}}, \bibinfo {author} {\bibfnamefont
  {S.}~\bibnamefont {Jain}}, \bibinfo {author} {\bibfnamefont {A.~O.}\
  \bibnamefont {Adeyeye}}, \ and\ \bibinfo {author} {\bibfnamefont
  {N.}~\bibnamefont {Singh}},\ }\href {\doibase 10.1103/PhysRevLett.107.127204}
  {\bibfield  {journal} {\bibinfo  {journal} {Phys. Rev. Lett.}\ }\textbf
  {\bibinfo {volume} {107}},\ \bibinfo {pages} {127204} (\bibinfo {year}
  {2011})}\BibitemShut {NoStop}%
\bibitem [{\citenamefont {Tacchi}\ \emph {et~al.}(2012)\citenamefont {Tacchi},
  \citenamefont {Duerr}, \citenamefont {Klos}, \citenamefont {Madami},
  \citenamefont {Neusser}, \citenamefont {Gubbiotti}, \citenamefont {Carlotti},
  \citenamefont {Krawczyk},\ and\ \citenamefont {Grundler}}]{Tacchi12}%
  \BibitemOpen
  \bibfield  {author} {\bibinfo {author} {\bibfnamefont {S.}~\bibnamefont
  {Tacchi}}, \bibinfo {author} {\bibfnamefont {G.}~\bibnamefont {Duerr}},
  \bibinfo {author} {\bibfnamefont {J.~W.}\ \bibnamefont {Klos}}, \bibinfo
  {author} {\bibfnamefont {M.}~\bibnamefont {Madami}}, \bibinfo {author}
  {\bibfnamefont {S.}~\bibnamefont {Neusser}}, \bibinfo {author} {\bibfnamefont
  {G.}~\bibnamefont {Gubbiotti}}, \bibinfo {author} {\bibfnamefont
  {G.}~\bibnamefont {Carlotti}}, \bibinfo {author} {\bibfnamefont
  {M.}~\bibnamefont {Krawczyk}}, \ and\ \bibinfo {author} {\bibfnamefont
  {D.}~\bibnamefont {Grundler}},\ }\href {\doibase
  10.1103/PhysRevLett.109.137202} {\bibfield  {journal} {\bibinfo  {journal}
  {Phys. Rev. Lett.}\ }\textbf {\bibinfo {volume} {109}},\ \bibinfo {pages}
  {137202} (\bibinfo {year} {2012})}\BibitemShut {NoStop}%
\bibitem [{\citenamefont {Chumak}\ \emph {et~al.}(2012)\citenamefont {Chumak},
  \citenamefont {Vasyuchka}, \citenamefont {Serga}, \citenamefont {Kostylev},
  \citenamefont {Tiberkevich},\ and\ \citenamefont {Hillebrands}}]{Chumak12}%
  \BibitemOpen
  \bibfield  {author} {\bibinfo {author} {\bibfnamefont {A.~V.}\ \bibnamefont
  {Chumak}}, \bibinfo {author} {\bibfnamefont {V.~I.}\ \bibnamefont
  {Vasyuchka}}, \bibinfo {author} {\bibfnamefont {A.~A.}\ \bibnamefont
  {Serga}}, \bibinfo {author} {\bibfnamefont {M.~P.}\ \bibnamefont {Kostylev}},
  \bibinfo {author} {\bibfnamefont {V.~S.}\ \bibnamefont {Tiberkevich}}, \ and\
  \bibinfo {author} {\bibfnamefont {B.}~\bibnamefont {Hillebrands}},\ }\href
  {http://link.aps.org/doi/10.1103/PhysRevLett.108.257207} {\bibfield
  {journal} {\bibinfo  {journal} {Phys. Rev. Lett.}\ }\textbf {\bibinfo
  {volume} {108}},\ \bibinfo {pages} {257207} (\bibinfo {year}
  {2012})}\BibitemShut {NoStop}%
\bibitem [{\citenamefont {Yu}\ \emph {et~al.}(2013)\citenamefont {Yu},
  \citenamefont {Duerr}, \citenamefont {Huber}, \citenamefont {Bahr},
  \citenamefont {Schwarze}, \citenamefont {Brandl},\ and\ \citenamefont
  {Grundler}}]{Yu13}%
  \BibitemOpen
  \bibfield  {author} {\bibinfo {author} {\bibfnamefont {H.}~\bibnamefont
  {Yu}}, \bibinfo {author} {\bibfnamefont {G.}~\bibnamefont {Duerr}}, \bibinfo
  {author} {\bibfnamefont {R.}~\bibnamefont {Huber}}, \bibinfo {author}
  {\bibfnamefont {M.}~\bibnamefont {Bahr}}, \bibinfo {author} {\bibfnamefont
  {T.}~\bibnamefont {Schwarze}}, \bibinfo {author} {\bibfnamefont
  {F.}~\bibnamefont {Brandl}}, \ and\ \bibinfo {author} {\bibfnamefont
  {D.}~\bibnamefont {Grundler}},\ }\href {http://dx.doi.org/10.1038/ncomms3702}
  {\bibfield  {journal} {\bibinfo  {journal} {Nat. Commun.}\ }\textbf {\bibinfo
  {volume} {4}},\ \bibinfo {pages} {2702} (\bibinfo {year} {2013})}\BibitemShut
  {NoStop}%
\bibitem [{\citenamefont {Krawczyk}\ and\ \citenamefont
  {Grundler}(2014)}]{Krawczyk14}%
  \BibitemOpen
  \bibfield  {author} {\bibinfo {author} {\bibfnamefont {M.}~\bibnamefont
  {Krawczyk}}\ and\ \bibinfo {author} {\bibfnamefont {D.}~\bibnamefont
  {Grundler}},\ }\href {http://stacks.iop.org/0953-8984/26/i=12/a=123202}
  {\bibfield  {journal} {\bibinfo  {journal} {J. Phys.: Condens. Matter}\
  }\textbf {\bibinfo {volume} {26}},\ \bibinfo {pages} {123202} (\bibinfo
  {year} {2014})}\BibitemShut {NoStop}%
\bibitem [{\citenamefont {K\"orner}\ \emph {et~al.}(2013)\citenamefont
  {K\"orner}, \citenamefont {Lenz}, \citenamefont {Gallardo}, \citenamefont
  {Fritzsche}, \citenamefont {M\"ucklich}, \citenamefont {Facsko},
  \citenamefont {Lindner}, \citenamefont {Landeros},\ and\ \citenamefont
  {Fassbender}}]{Korner13}%
  \BibitemOpen
  \bibfield  {author} {\bibinfo {author} {\bibfnamefont {M.}~\bibnamefont
  {K\"orner}}, \bibinfo {author} {\bibfnamefont {K.}~\bibnamefont {Lenz}},
  \bibinfo {author} {\bibfnamefont {R.~A.}\ \bibnamefont {Gallardo}}, \bibinfo
  {author} {\bibfnamefont {M.}~\bibnamefont {Fritzsche}}, \bibinfo {author}
  {\bibfnamefont {A.}~\bibnamefont {M\"ucklich}}, \bibinfo {author}
  {\bibfnamefont {S.}~\bibnamefont {Facsko}}, \bibinfo {author} {\bibfnamefont
  {J.}~\bibnamefont {Lindner}}, \bibinfo {author} {\bibfnamefont
  {P.}~\bibnamefont {Landeros}}, \ and\ \bibinfo {author} {\bibfnamefont
  {J.}~\bibnamefont {Fassbender}},\ }\href {\doibase
  10.1103/PhysRevB.88.054405} {\bibfield  {journal} {\bibinfo  {journal} {Phys.
  Rev. B}\ }\textbf {\bibinfo {volume} {88}},\ \bibinfo {pages} {054405}
  (\bibinfo {year} {2013})}\BibitemShut {NoStop}%
\bibitem [{\citenamefont {Gallardo}\ \emph {et~al.}(2014)\citenamefont
  {Gallardo}, \citenamefont {Banholzer}, \citenamefont {Wagner}, \citenamefont
  {K{\"o}rner}, \citenamefont {Lenz}, \citenamefont {Farle}, \citenamefont
  {Lindner}, \citenamefont {Fassbender},\ and\ \citenamefont
  {Landeros}}]{Gallardo14}%
  \BibitemOpen
  \bibfield  {author} {\bibinfo {author} {\bibfnamefont {R.~A.}\ \bibnamefont
  {Gallardo}}, \bibinfo {author} {\bibfnamefont {A.}~\bibnamefont {Banholzer}},
  \bibinfo {author} {\bibfnamefont {K.}~\bibnamefont {Wagner}}, \bibinfo
  {author} {\bibfnamefont {M.}~\bibnamefont {K{\"o}rner}}, \bibinfo {author}
  {\bibfnamefont {K.}~\bibnamefont {Lenz}}, \bibinfo {author} {\bibfnamefont
  {M.}~\bibnamefont {Farle}}, \bibinfo {author} {\bibfnamefont
  {J.}~\bibnamefont {Lindner}}, \bibinfo {author} {\bibfnamefont
  {J.}~\bibnamefont {Fassbender}}, \ and\ \bibinfo {author} {\bibfnamefont
  {P.}~\bibnamefont {Landeros}},\ }\href
  {http://stacks.iop.org/1367-2630/16/i=2/a=023015} {\bibfield  {journal}
  {\bibinfo  {journal} {New J. Phys.}\ }\textbf {\bibinfo {volume} {16}},\
  \bibinfo {pages} {023015} (\bibinfo {year} {2014})}\BibitemShut {NoStop}%
\bibitem [{\citenamefont {Sebastian}\ \emph {et~al.}(2015)\citenamefont
  {Sebastian}, \citenamefont {Schultheiss}, \citenamefont {Obry}, \citenamefont
  {Hillebrands},\ and\ \citenamefont {Schultheiss}}]{Sebastian15}%
  \BibitemOpen
  \bibfield  {author} {\bibinfo {author} {\bibfnamefont {T.}~\bibnamefont
  {Sebastian}}, \bibinfo {author} {\bibfnamefont {K.}~\bibnamefont
  {Schultheiss}}, \bibinfo {author} {\bibfnamefont {B.}~\bibnamefont {Obry}},
  \bibinfo {author} {\bibfnamefont {B.}~\bibnamefont {Hillebrands}}, \ and\
  \bibinfo {author} {\bibfnamefont {H.}~\bibnamefont {Schultheiss}},\ }\href
  {\doibase 10.3389/fphy.2015.00035} {\bibfield  {journal} {\bibinfo  {journal}
  {Front. Phys.}\ }\textbf {\bibinfo {volume} {3}},\ \bibinfo {pages} {35}
  (\bibinfo {year} {2015})}\BibitemShut {NoStop}%
\bibitem [{\citenamefont {Ciubotaru}\ \emph {et~al.}(2012)\citenamefont
  {Ciubotaru}, \citenamefont {Chumak}, \citenamefont {Grigoryeva},
  \citenamefont {Serga},\ and\ \citenamefont {Hillebrands}}]{Ciubotaru12}%
  \BibitemOpen
  \bibfield  {author} {\bibinfo {author} {\bibfnamefont {F.}~\bibnamefont
  {Ciubotaru}}, \bibinfo {author} {\bibfnamefont {A.~V.}\ \bibnamefont
  {Chumak}}, \bibinfo {author} {\bibfnamefont {N.~Y.}\ \bibnamefont
  {Grigoryeva}}, \bibinfo {author} {\bibfnamefont {A.~A.}\ \bibnamefont
  {Serga}}, \ and\ \bibinfo {author} {\bibfnamefont {B.}~\bibnamefont
  {Hillebrands}},\ }\href {http://stacks.iop.org/0022-3727/45/i=25/a=255002}
  {\bibfield  {journal} {\bibinfo  {journal} {J. Phys. D: Appl. Phys.}\
  }\textbf {\bibinfo {volume} {45}},\ \bibinfo {pages} {255002} (\bibinfo
  {year} {2012})}\BibitemShut {NoStop}%
\bibitem [{\citenamefont {K\l{}os}\ \emph {et~al.}(2012)\citenamefont
  {K\l{}os}, \citenamefont {Kumar}, \citenamefont {Romero-Vivas}, \citenamefont
  {Fangohr}, \citenamefont {Franchin}, \citenamefont {Krawczyk},\ and\
  \citenamefont {Barman}}]{Klos12}%
  \BibitemOpen
  \bibfield  {author} {\bibinfo {author} {\bibfnamefont {J.~W.}\ \bibnamefont
  {K\l{}os}}, \bibinfo {author} {\bibfnamefont {D.}~\bibnamefont {Kumar}},
  \bibinfo {author} {\bibfnamefont {J.}~\bibnamefont {Romero-Vivas}}, \bibinfo
  {author} {\bibfnamefont {H.}~\bibnamefont {Fangohr}}, \bibinfo {author}
  {\bibfnamefont {M.}~\bibnamefont {Franchin}}, \bibinfo {author}
  {\bibfnamefont {M.}~\bibnamefont {Krawczyk}}, \ and\ \bibinfo {author}
  {\bibfnamefont {A.}~\bibnamefont {Barman}},\ }\href
  {http://link.aps.org/doi/10.1103/PhysRevB.86.184433} {\bibfield  {journal}
  {\bibinfo  {journal} {Phys. Rev. B}\ }\textbf {\bibinfo {volume} {86}},\
  \bibinfo {pages} {184433} (\bibinfo {year} {2012})}\BibitemShut {NoStop}%
\bibitem [{\citenamefont {Krawczyk}\ \emph {et~al.}(2013)\citenamefont
  {Krawczyk}, \citenamefont {Mamica}, \citenamefont {Mruczkiewicz},
  \citenamefont {Klos}, \citenamefont {Tacchi}, \citenamefont {Madami},
  \citenamefont {Gubbiotti}, \citenamefont {Duerr},\ and\ \citenamefont
  {Grundler}}]{Krawczyk13}%
  \BibitemOpen
  \bibfield  {author} {\bibinfo {author} {\bibfnamefont {M.}~\bibnamefont
  {Krawczyk}}, \bibinfo {author} {\bibfnamefont {S.}~\bibnamefont {Mamica}},
  \bibinfo {author} {\bibfnamefont {M.}~\bibnamefont {Mruczkiewicz}}, \bibinfo
  {author} {\bibfnamefont {J.~W.}\ \bibnamefont {Klos}}, \bibinfo {author}
  {\bibfnamefont {S.}~\bibnamefont {Tacchi}}, \bibinfo {author} {\bibfnamefont
  {M.}~\bibnamefont {Madami}}, \bibinfo {author} {\bibfnamefont
  {G.}~\bibnamefont {Gubbiotti}}, \bibinfo {author} {\bibfnamefont
  {G.}~\bibnamefont {Duerr}}, \ and\ \bibinfo {author} {\bibfnamefont
  {D.}~\bibnamefont {Grundler}},\ }\href
  {http://stacks.iop.org/0022-3727/46/i=49/a=495003} {\bibfield  {journal}
  {\bibinfo  {journal} {J. Phys. D: Appl. Phys.}\ }\textbf {\bibinfo {volume}
  {46}},\ \bibinfo {pages} {495003} (\bibinfo {year} {2013})}\BibitemShut
  {NoStop}%
\bibitem [{\citenamefont {Gubbiotti}\ \emph {et~al.}(2014)\citenamefont
  {Gubbiotti}, \citenamefont {Kostylev}, \citenamefont {Tacchi}, \citenamefont
  {Madami}, \citenamefont {Carlotti}, \citenamefont {Ding}, \citenamefont
  {Adeyeye}, \citenamefont {Zighem}, \citenamefont {Stashkevich}, \citenamefont
  {Ivanov},\ and\ \citenamefont {Samarin}}]{Gubbiotti14}%
  \BibitemOpen
  \bibfield  {author} {\bibinfo {author} {\bibfnamefont {G.}~\bibnamefont
  {Gubbiotti}}, \bibinfo {author} {\bibfnamefont {M.}~\bibnamefont {Kostylev}},
  \bibinfo {author} {\bibfnamefont {S.}~\bibnamefont {Tacchi}}, \bibinfo
  {author} {\bibfnamefont {M.}~\bibnamefont {Madami}}, \bibinfo {author}
  {\bibfnamefont {G.}~\bibnamefont {Carlotti}}, \bibinfo {author}
  {\bibfnamefont {J.}~\bibnamefont {Ding}}, \bibinfo {author} {\bibfnamefont
  {A.~O.}\ \bibnamefont {Adeyeye}}, \bibinfo {author} {\bibfnamefont
  {F.}~\bibnamefont {Zighem}}, \bibinfo {author} {\bibfnamefont {A.~A.}\
  \bibnamefont {Stashkevich}}, \bibinfo {author} {\bibfnamefont
  {E.}~\bibnamefont {Ivanov}}, \ and\ \bibinfo {author} {\bibfnamefont
  {S.}~\bibnamefont {Samarin}},\ }\href
  {http://stacks.iop.org/0022-3727/47/i=10/a=105003} {\bibfield  {journal}
  {\bibinfo  {journal} {J. Phys. D: Appl. Phys.}\ }\textbf {\bibinfo {volume}
  {47}},\ \bibinfo {pages} {105003} (\bibinfo {year} {2014})}\BibitemShut
  {NoStop}%
\bibitem [{\citenamefont {Barsukov}\ \emph {et~al.}(2011)\citenamefont
  {Barsukov}, \citenamefont {R\"omer}, \citenamefont {Meckenstock},
  \citenamefont {Lenz}, \citenamefont {Lindner}, \citenamefont {Hemken~to
  Krax}, \citenamefont {Banholzer}, \citenamefont {K\"orner}, \citenamefont
  {Grebing}, \citenamefont {Fassbender},\ and\ \citenamefont
  {Farle}}]{Barsukov11}%
  \BibitemOpen
  \bibfield  {author} {\bibinfo {author} {\bibfnamefont {I.}~\bibnamefont
  {Barsukov}}, \bibinfo {author} {\bibfnamefont {F.~M.}\ \bibnamefont
  {R\"omer}}, \bibinfo {author} {\bibfnamefont {R.}~\bibnamefont
  {Meckenstock}}, \bibinfo {author} {\bibfnamefont {K.}~\bibnamefont {Lenz}},
  \bibinfo {author} {\bibfnamefont {J.}~\bibnamefont {Lindner}}, \bibinfo
  {author} {\bibfnamefont {S.}~\bibnamefont {Hemken~to Krax}}, \bibinfo
  {author} {\bibfnamefont {A.}~\bibnamefont {Banholzer}}, \bibinfo {author}
  {\bibfnamefont {M.}~\bibnamefont {K\"orner}}, \bibinfo {author}
  {\bibfnamefont {J.}~\bibnamefont {Grebing}}, \bibinfo {author} {\bibfnamefont
  {J.}~\bibnamefont {Fassbender}}, \ and\ \bibinfo {author} {\bibfnamefont
  {M.}~\bibnamefont {Farle}},\ }\href {\doibase 10.1103/PhysRevB.84.140410}
  {\bibfield  {journal} {\bibinfo  {journal} {Phys. Rev. B}\ }\textbf {\bibinfo
  {volume} {84}},\ \bibinfo {pages} {140410} (\bibinfo {year}
  {2011})}\BibitemShut {NoStop}%
\bibitem [{\citenamefont {Obry}\ \emph {et~al.}(2013)\citenamefont {Obry},
  \citenamefont {Pirro}, \citenamefont {Br{\"a}cher}, \citenamefont {Chumak},
  \citenamefont {Osten}, \citenamefont {Ciubotaru}, \citenamefont {Serga},
  \citenamefont {Fassbender},\ and\ \citenamefont {Hillebrands}}]{Obry13}%
  \BibitemOpen
  \bibfield  {author} {\bibinfo {author} {\bibfnamefont {B.}~\bibnamefont
  {Obry}}, \bibinfo {author} {\bibfnamefont {P.}~\bibnamefont {Pirro}},
  \bibinfo {author} {\bibfnamefont {T.}~\bibnamefont {Br{\"a}cher}}, \bibinfo
  {author} {\bibfnamefont {A.~V.}\ \bibnamefont {Chumak}}, \bibinfo {author}
  {\bibfnamefont {J.}~\bibnamefont {Osten}}, \bibinfo {author} {\bibfnamefont
  {F.}~\bibnamefont {Ciubotaru}}, \bibinfo {author} {\bibfnamefont {A.~A.}\
  \bibnamefont {Serga}}, \bibinfo {author} {\bibfnamefont {J.}~\bibnamefont
  {Fassbender}}, \ and\ \bibinfo {author} {\bibfnamefont {B.}~\bibnamefont
  {Hillebrands}},\ }\href
  {http://scitation.aip.org/content/aip/journal/apl/102/20/10.1063/1.4807721}
  {\bibfield  {journal} {\bibinfo  {journal} {Appl. Phys. Lett.}\ }\textbf
  {\bibinfo {volume} {102}},\ \bibinfo {eid} {202403} (\bibinfo {year}
  {2013})}\BibitemShut {NoStop}%
\bibitem [{\citenamefont {Landeros}\ and\ \citenamefont
  {Mills}(2012)}]{Landeros12}%
  \BibitemOpen
  \bibfield  {author} {\bibinfo {author} {\bibfnamefont {P.}~\bibnamefont
  {Landeros}}\ and\ \bibinfo {author} {\bibfnamefont {D.~L.}\ \bibnamefont
  {Mills}},\ }\href {\doibase 10.1103/PhysRevB.85.054424} {\bibfield  {journal}
  {\bibinfo  {journal} {Phys. Rev. B}\ }\textbf {\bibinfo {volume} {85}},\
  \bibinfo {pages} {054424} (\bibinfo {year} {2012})}\BibitemShut {NoStop}%
\bibitem [{\citenamefont {Chumak}\ \emph
  {et~al.}(2009{\natexlab{b}})\citenamefont {Chumak}, \citenamefont {Neumann},
  \citenamefont {Serga}, \citenamefont {Hillebrands},\ and\ \citenamefont
  {Kostylev}}]{Chumak09b}%
  \BibitemOpen
  \bibfield  {author} {\bibinfo {author} {\bibfnamefont {A.~V.}\ \bibnamefont
  {Chumak}}, \bibinfo {author} {\bibfnamefont {T.}~\bibnamefont {Neumann}},
  \bibinfo {author} {\bibfnamefont {A.~A.}\ \bibnamefont {Serga}}, \bibinfo
  {author} {\bibfnamefont {B.}~\bibnamefont {Hillebrands}}, \ and\ \bibinfo
  {author} {\bibfnamefont {M.~P.}\ \bibnamefont {Kostylev}},\ }\href
  {http://stacks.iop.org/0022-3727/42/i=20/a=205005} {\bibfield  {journal}
  {\bibinfo  {journal} {J. Phys. D: Appl. Phys.}\ }\textbf {\bibinfo {volume}
  {42}},\ \bibinfo {pages} {205005} (\bibinfo {year}
  {2009}{\natexlab{b}})}\BibitemShut {NoStop}%
\bibitem [{\citenamefont {Chumak}\ \emph {et~al.}(2010)\citenamefont {Chumak},
  \citenamefont {Tiberkevich}, \citenamefont {Karenowska}, \citenamefont
  {Serga}, \citenamefont {Gregg}, \citenamefont {Slavin},\ and\ \citenamefont
  {Hillebrands}}]{Chumak14b}%
  \BibitemOpen
  \bibfield  {author} {\bibinfo {author} {\bibfnamefont {A.~V.}\ \bibnamefont
  {Chumak}}, \bibinfo {author} {\bibfnamefont {V.~S.}\ \bibnamefont
  {Tiberkevich}}, \bibinfo {author} {\bibfnamefont {A.~D.}\ \bibnamefont
  {Karenowska}}, \bibinfo {author} {\bibfnamefont {A.~A.}\ \bibnamefont
  {Serga}}, \bibinfo {author} {\bibfnamefont {J.~F.}\ \bibnamefont {Gregg}},
  \bibinfo {author} {\bibfnamefont {A.~N.}\ \bibnamefont {Slavin}}, \ and\
  \bibinfo {author} {\bibfnamefont {B.}~\bibnamefont {Hillebrands}},\ }\href
  {http://dx.doi.org/10.1038/ncomms1142} {\bibfield  {journal} {\bibinfo
  {journal} {Nat. Commun.}\ }\textbf {\bibinfo {volume} {1}},\ \bibinfo {pages}
  {141} (\bibinfo {year} {2010})}\BibitemShut {NoStop}%
\bibitem [{\citenamefont {Vogel}\ \emph {et~al.}(2015)\citenamefont {Vogel},
  \citenamefont {Chumak}, \citenamefont {Waller}, \citenamefont {Langner},
  \citenamefont {Vasyuchka}, \citenamefont {Hillebrands},\ and\ \citenamefont
  {von Freymann}}]{Vogel15}%
  \BibitemOpen
  \bibfield  {author} {\bibinfo {author} {\bibfnamefont {M.}~\bibnamefont
  {Vogel}}, \bibinfo {author} {\bibfnamefont {A.~V.}\ \bibnamefont {Chumak}},
  \bibinfo {author} {\bibfnamefont {E.~H.}\ \bibnamefont {Waller}}, \bibinfo
  {author} {\bibfnamefont {T.}~\bibnamefont {Langner}}, \bibinfo {author}
  {\bibfnamefont {V.~I.}\ \bibnamefont {Vasyuchka}}, \bibinfo {author}
  {\bibfnamefont {B.}~\bibnamefont {Hillebrands}}, \ and\ \bibinfo {author}
  {\bibfnamefont {G.}~\bibnamefont {von Freymann}},\ }\href
  {http://dx.doi.org/10.1038/nphys3325} {\bibfield  {journal} {\bibinfo
  {journal} {Nat. Phys.}\ }\textbf {\bibinfo {volume} {11}},\ \bibinfo {pages}
  {487} (\bibinfo {year} {2015})}\BibitemShut {NoStop}%
\bibitem [{\citenamefont {Grundler}(2015)}]{Grundler15}%
  \BibitemOpen
  \bibfield  {author} {\bibinfo {author} {\bibfnamefont {D.}~\bibnamefont
  {Grundler}},\ }\href {http://dx.doi.org/10.1038/nphys3349} {\bibfield
  {journal} {\bibinfo  {journal} {Nat. Phys.}\ }\textbf {\bibinfo {volume}
  {11}},\ \bibinfo {pages} {438} (\bibinfo {year} {2015})}\BibitemShut
  {NoStop}%
\bibitem [{\citenamefont {Lin}\ \emph {et~al.}(2011)\citenamefont {Lin},
  \citenamefont {Lim}, \citenamefont {Wang}, \citenamefont {Ng},\ and\
  \citenamefont {Kuok}}]{Lin11}%
  \BibitemOpen
  \bibfield  {author} {\bibinfo {author} {\bibfnamefont {C.~S.}\ \bibnamefont
  {Lin}}, \bibinfo {author} {\bibfnamefont {H.~S.}\ \bibnamefont {Lim}},
  \bibinfo {author} {\bibfnamefont {Z.~K.}\ \bibnamefont {Wang}}, \bibinfo
  {author} {\bibfnamefont {S.~C.}\ \bibnamefont {Ng}}, \ and\ \bibinfo {author}
  {\bibfnamefont {M.~H.}\ \bibnamefont {Kuok}},\ }\href
  {http://scitation.aip.org/content/aip/journal/apl/98/2/10.1063/1.3541886}
  {\bibfield  {journal} {\bibinfo  {journal} {Appl. Phys. Lett.}\ }\textbf
  {\bibinfo {volume} {98}},\ \bibinfo {eid} {022504} (\bibinfo {year}
  {2011})}\BibitemShut {NoStop}%
\bibitem [{\citenamefont {Sokolovskyy}\ and\ \citenamefont
  {Krawczyk}(2011)}]{Sokolovskyy11}%
  \BibitemOpen
  \bibfield  {author} {\bibinfo {author} {\bibfnamefont {M.}~\bibnamefont
  {Sokolovskyy}}\ and\ \bibinfo {author} {\bibfnamefont {M.}~\bibnamefont
  {Krawczyk}},\ }\href {\doibase 10.1007/s11051-011-0303-5} {\bibfield
  {journal} {\bibinfo  {journal} {J. Nanopart. Res.}\ }\textbf {\bibinfo
  {volume} {13}},\ \bibinfo {pages} {6085} (\bibinfo {year}
  {2011})}\BibitemShut {NoStop}%
\bibitem [{\citenamefont {Mruczkiewicz}\ \emph {et~al.}(2013)\citenamefont
  {Mruczkiewicz}, \citenamefont {Krawczyk}, \citenamefont {Sakharov},
  \citenamefont {Khivintsev}, \citenamefont {Filimonov},\ and\ \citenamefont
  {Nikitov}}]{Mruczkiewicz13}%
  \BibitemOpen
  \bibfield  {author} {\bibinfo {author} {\bibfnamefont {M.}~\bibnamefont
  {Mruczkiewicz}}, \bibinfo {author} {\bibfnamefont {M.}~\bibnamefont
  {Krawczyk}}, \bibinfo {author} {\bibfnamefont {V.~K.}\ \bibnamefont
  {Sakharov}}, \bibinfo {author} {\bibfnamefont {Y.~V.}\ \bibnamefont
  {Khivintsev}}, \bibinfo {author} {\bibfnamefont {Y.~A.}\ \bibnamefont
  {Filimonov}}, \ and\ \bibinfo {author} {\bibfnamefont {S.~A.}\ \bibnamefont
  {Nikitov}},\ }\href
  {http://scitation.aip.org/content/aip/journal/jap/113/9/10.1063/1.4793085}
  {\bibfield  {journal} {\bibinfo  {journal} {J. Appl. Phys.}\ }\textbf
  {\bibinfo {volume} {113}},\ \bibinfo {eid} {093908} (\bibinfo {year}
  {2013})}\BibitemShut {NoStop}%
\bibitem [{\citenamefont {Mathieu}\ \emph {et~al.}(1998)\citenamefont
  {Mathieu}, \citenamefont {Jorzick}, \citenamefont {Frank}, \citenamefont
  {Demokritov}, \citenamefont {Slavin}, \citenamefont {Hillebrands},
  \citenamefont {Bartenlian}, \citenamefont {Chappert}, \citenamefont
  {Decanini}, \citenamefont {Rousseaux},\ and\ \citenamefont
  {Cambril}}]{Mathieu98}%
  \BibitemOpen
  \bibfield  {author} {\bibinfo {author} {\bibfnamefont {C.}~\bibnamefont
  {Mathieu}}, \bibinfo {author} {\bibfnamefont {J.}~\bibnamefont {Jorzick}},
  \bibinfo {author} {\bibfnamefont {A.}~\bibnamefont {Frank}}, \bibinfo
  {author} {\bibfnamefont {S.~O.}\ \bibnamefont {Demokritov}}, \bibinfo
  {author} {\bibfnamefont {A.~N.}\ \bibnamefont {Slavin}}, \bibinfo {author}
  {\bibfnamefont {B.}~\bibnamefont {Hillebrands}}, \bibinfo {author}
  {\bibfnamefont {B.}~\bibnamefont {Bartenlian}}, \bibinfo {author}
  {\bibfnamefont {C.}~\bibnamefont {Chappert}}, \bibinfo {author}
  {\bibfnamefont {D.}~\bibnamefont {Decanini}}, \bibinfo {author}
  {\bibfnamefont {F.}~\bibnamefont {Rousseaux}}, \ and\ \bibinfo {author}
  {\bibfnamefont {E.}~\bibnamefont {Cambril}},\ }\href {\doibase
  10.1103/PhysRevLett.81.3968} {\bibfield  {journal} {\bibinfo  {journal}
  {Phys. Rev. Lett.}\ }\textbf {\bibinfo {volume} {81}},\ \bibinfo {pages}
  {3968} (\bibinfo {year} {1998})}\BibitemShut {NoStop}%
\bibitem [{\citenamefont {Jorzick}\ \emph {et~al.}(1999)\citenamefont
  {Jorzick}, \citenamefont {Demokritov}, \citenamefont {Mathieu}, \citenamefont
  {Hillebrands}, \citenamefont {Bartenlian}, \citenamefont {Chappert},
  \citenamefont {Rousseaux},\ and\ \citenamefont {Slavin}}]{Jorzick99}%
  \BibitemOpen
  \bibfield  {author} {\bibinfo {author} {\bibfnamefont {J.}~\bibnamefont
  {Jorzick}}, \bibinfo {author} {\bibfnamefont {S.~O.}\ \bibnamefont
  {Demokritov}}, \bibinfo {author} {\bibfnamefont {C.}~\bibnamefont {Mathieu}},
  \bibinfo {author} {\bibfnamefont {B.}~\bibnamefont {Hillebrands}}, \bibinfo
  {author} {\bibfnamefont {B.}~\bibnamefont {Bartenlian}}, \bibinfo {author}
  {\bibfnamefont {C.}~\bibnamefont {Chappert}}, \bibinfo {author}
  {\bibfnamefont {F.}~\bibnamefont {Rousseaux}}, \ and\ \bibinfo {author}
  {\bibfnamefont {A.~N.}\ \bibnamefont {Slavin}},\ }\href {\doibase
  10.1103/PhysRevB.60.15194} {\bibfield  {journal} {\bibinfo  {journal} {Phys.
  Rev. B}\ }\textbf {\bibinfo {volume} {60}},\ \bibinfo {pages} {15194}
  (\bibinfo {year} {1999})}\BibitemShut {NoStop}%
\bibitem [{\citenamefont {Di}\ \emph {et~al.}(2015{\natexlab{a}})\citenamefont
  {Di}, \citenamefont {Feng}, \citenamefont {Piramanayagam}, \citenamefont
  {Zhang}, \citenamefont {Lim}, \citenamefont {Ng},\ and\ \citenamefont
  {Kuok}}]{Dib15}%
  \BibitemOpen
  \bibfield  {author} {\bibinfo {author} {\bibfnamefont {K.}~\bibnamefont
  {Di}}, \bibinfo {author} {\bibfnamefont {S.~X.}\ \bibnamefont {Feng}},
  \bibinfo {author} {\bibfnamefont {S.~N.}\ \bibnamefont {Piramanayagam}},
  \bibinfo {author} {\bibfnamefont {V.~L.}\ \bibnamefont {Zhang}}, \bibinfo
  {author} {\bibfnamefont {H.~S.}\ \bibnamefont {Lim}}, \bibinfo {author}
  {\bibfnamefont {S.~C.}\ \bibnamefont {Ng}}, \ and\ \bibinfo {author}
  {\bibfnamefont {M.~H.}\ \bibnamefont {Kuok}},\ }\href
  {http://dx.doi.org/10.1038/srep10153} {\bibfield  {journal} {\bibinfo
  {journal} {Sci. Rep.}\ }\textbf {\bibinfo {volume} {5}},\ \bibinfo {pages}
  {10153 EP } (\bibinfo {year} {2015}{\natexlab{a}})}\BibitemShut {NoStop}%
\bibitem [{\citenamefont {Gladii}\ \emph {et~al.}(2016)\citenamefont {Gladii},
  \citenamefont {Haidar}, \citenamefont {Henry}, \citenamefont {Kostylev},\
  and\ \citenamefont {Bailleul}}]{Gladii16}%
  \BibitemOpen
  \bibfield  {author} {\bibinfo {author} {\bibfnamefont {O.}~\bibnamefont
  {Gladii}}, \bibinfo {author} {\bibfnamefont {M.}~\bibnamefont {Haidar}},
  \bibinfo {author} {\bibfnamefont {Y.}~\bibnamefont {Henry}}, \bibinfo
  {author} {\bibfnamefont {M.}~\bibnamefont {Kostylev}}, \ and\ \bibinfo
  {author} {\bibfnamefont {M.}~\bibnamefont {Bailleul}},\ }\href {\doibase
  10.1103/PhysRevB.93.054430} {\bibfield  {journal} {\bibinfo  {journal} {Phys.
  Rev. B}\ }\textbf {\bibinfo {volume} {93}},\ \bibinfo {pages} {054430}
  (\bibinfo {year} {2016})}\BibitemShut {NoStop}%
\bibitem [{\citenamefont {Zakeri}\ \emph {et~al.}(2010)\citenamefont {Zakeri},
  \citenamefont {Zhang}, \citenamefont {Prokop}, \citenamefont {Chuang},
  \citenamefont {Sakr}, \citenamefont {Tang},\ and\ \citenamefont
  {Kirschner}}]{Zakeri10}%
  \BibitemOpen
  \bibfield  {author} {\bibinfo {author} {\bibfnamefont {K.}~\bibnamefont
  {Zakeri}}, \bibinfo {author} {\bibfnamefont {Y.}~\bibnamefont {Zhang}},
  \bibinfo {author} {\bibfnamefont {J.}~\bibnamefont {Prokop}}, \bibinfo
  {author} {\bibfnamefont {T.-H.}\ \bibnamefont {Chuang}}, \bibinfo {author}
  {\bibfnamefont {N.}~\bibnamefont {Sakr}}, \bibinfo {author} {\bibfnamefont
  {W.~X.}\ \bibnamefont {Tang}}, \ and\ \bibinfo {author} {\bibfnamefont
  {J.}~\bibnamefont {Kirschner}},\ }\href {\doibase
  10.1103/PhysRevLett.104.137203} {\bibfield  {journal} {\bibinfo  {journal}
  {Phys. Rev. Lett.}\ }\textbf {\bibinfo {volume} {104}},\ \bibinfo {pages}
  {137203} (\bibinfo {year} {2010})}\BibitemShut {NoStop}%
\bibitem [{\citenamefont {Cort{\'e}s-Ortu{\~n}o}\ and\ \citenamefont
  {Landeros}(2013)}]{Cortes13}%
  \BibitemOpen
  \bibfield  {author} {\bibinfo {author} {\bibfnamefont {D.}~\bibnamefont
  {Cort{\'e}s-Ortu{\~n}o}}\ and\ \bibinfo {author} {\bibfnamefont
  {P.}~\bibnamefont {Landeros}},\ }\href
  {http://stacks.iop.org/0953-8984/25/i=15/a=156001} {\bibfield  {journal}
  {\bibinfo  {journal} {J. Phys.: Condens. Matter}\ }\textbf {\bibinfo {volume}
  {25}},\ \bibinfo {pages} {156001} (\bibinfo {year} {2013})}\BibitemShut
  {NoStop}%
\bibitem [{\citenamefont {Di}\ \emph {et~al.}(2015{\natexlab{b}})\citenamefont
  {Di}, \citenamefont {Zhang}, \citenamefont {Lim}, \citenamefont {Ng},
  \citenamefont {Kuok}, \citenamefont {Yu}, \citenamefont {Yoon}, \citenamefont
  {Qiu},\ and\ \citenamefont {Yang}}]{Kai15}%
  \BibitemOpen
  \bibfield  {author} {\bibinfo {author} {\bibfnamefont {K.}~\bibnamefont
  {Di}}, \bibinfo {author} {\bibfnamefont {V.~L.}\ \bibnamefont {Zhang}},
  \bibinfo {author} {\bibfnamefont {H.~S.}\ \bibnamefont {Lim}}, \bibinfo
  {author} {\bibfnamefont {S.~C.}\ \bibnamefont {Ng}}, \bibinfo {author}
  {\bibfnamefont {M.~H.}\ \bibnamefont {Kuok}}, \bibinfo {author}
  {\bibfnamefont {J.}~\bibnamefont {Yu}}, \bibinfo {author} {\bibfnamefont
  {J.}~\bibnamefont {Yoon}}, \bibinfo {author} {\bibfnamefont {X.}~\bibnamefont
  {Qiu}}, \ and\ \bibinfo {author} {\bibfnamefont {H.}~\bibnamefont {Yang}},\
  }\href {\doibase 10.1103/PhysRevLett.114.047201} {\bibfield  {journal}
  {\bibinfo  {journal} {Phys. Rev. Lett.}\ }\textbf {\bibinfo {volume} {114}},\
  \bibinfo {pages} {047201} (\bibinfo {year} {2015}{\natexlab{b}})}\BibitemShut
  {NoStop}%
\bibitem [{\citenamefont {Zhang}\ \emph {et~al.}(2015)\citenamefont {Zhang},
  \citenamefont {Di}, \citenamefont {Lim}, \citenamefont {Ng}, \citenamefont
  {Kuok}, \citenamefont {Yu}, \citenamefont {Yoon}, \citenamefont {Qiu},\ and\
  \citenamefont {Yang}}]{Zhang15}%
  \BibitemOpen
  \bibfield  {author} {\bibinfo {author} {\bibfnamefont {V.~L.}\ \bibnamefont
  {Zhang}}, \bibinfo {author} {\bibfnamefont {K.}~\bibnamefont {Di}}, \bibinfo
  {author} {\bibfnamefont {H.~S.}\ \bibnamefont {Lim}}, \bibinfo {author}
  {\bibfnamefont {S.~C.}\ \bibnamefont {Ng}}, \bibinfo {author} {\bibfnamefont
  {M.~H.}\ \bibnamefont {Kuok}}, \bibinfo {author} {\bibfnamefont
  {J.}~\bibnamefont {Yu}}, \bibinfo {author} {\bibfnamefont {J.}~\bibnamefont
  {Yoon}}, \bibinfo {author} {\bibfnamefont {X.}~\bibnamefont {Qiu}}, \ and\
  \bibinfo {author} {\bibfnamefont {H.}~\bibnamefont {Yang}},\ }\href
  {http://scitation.aip.org/content/aip/journal/apl/107/2/10.1063/1.4926862}
  {\bibfield  {journal} {\bibinfo  {journal} {Appl. Phys. Lett.}\ }\textbf
  {\bibinfo {volume} {107}},\ \bibinfo {eid} {022402} (\bibinfo {year}
  {2015})}\BibitemShut {NoStop}%
\bibitem [{\citenamefont {Cho}\ \emph {et~al.}(2015)\citenamefont {Cho},
  \citenamefont {Kim}, \citenamefont {Lee}, \citenamefont {Kim}, \citenamefont
  {Lavrijsen}, \citenamefont {Solignac}, \citenamefont {Yin}, \citenamefont
  {Han}, \citenamefont {van Hoof}, \citenamefont {Swagten}, \citenamefont
  {Koopmans},\ and\ \citenamefont {You}}]{Cho15}%
  \BibitemOpen
  \bibfield  {author} {\bibinfo {author} {\bibfnamefont {J.}~\bibnamefont
  {Cho}}, \bibinfo {author} {\bibfnamefont {N.-H.}\ \bibnamefont {Kim}},
  \bibinfo {author} {\bibfnamefont {S.}~\bibnamefont {Lee}}, \bibinfo {author}
  {\bibfnamefont {J.-S.}\ \bibnamefont {Kim}}, \bibinfo {author} {\bibfnamefont
  {R.}~\bibnamefont {Lavrijsen}}, \bibinfo {author} {\bibfnamefont
  {A.}~\bibnamefont {Solignac}}, \bibinfo {author} {\bibfnamefont
  {Y.}~\bibnamefont {Yin}}, \bibinfo {author} {\bibfnamefont {D.-S.}\
  \bibnamefont {Han}}, \bibinfo {author} {\bibfnamefont {N.~J.~J.}\
  \bibnamefont {van Hoof}}, \bibinfo {author} {\bibfnamefont {H.~J.~M.}\
  \bibnamefont {Swagten}}, \bibinfo {author} {\bibfnamefont {B.}~\bibnamefont
  {Koopmans}}, \ and\ \bibinfo {author} {\bibfnamefont {C.-Y.}\ \bibnamefont
  {You}},\ }\href {http://dx.doi.org/10.1038/ncomms8635} {\bibfield  {journal}
  {\bibinfo  {journal} {Nat. Commun.}\ }\textbf {\bibinfo {volume} {6}}
  (\bibinfo {year} {2015})}\BibitemShut {NoStop}%
\bibitem [{\citenamefont {Nembach}\ \emph {et~al.}(2015)\citenamefont
  {Nembach}, \citenamefont {Shaw}, \citenamefont {Weiler}, \citenamefont
  {Jue},\ and\ \citenamefont {Silva}}]{Nembach15}%
  \BibitemOpen
  \bibfield  {author} {\bibinfo {author} {\bibfnamefont {H.~T.}\ \bibnamefont
  {Nembach}}, \bibinfo {author} {\bibfnamefont {J.~M.}\ \bibnamefont {Shaw}},
  \bibinfo {author} {\bibfnamefont {M.}~\bibnamefont {Weiler}}, \bibinfo
  {author} {\bibfnamefont {E.}~\bibnamefont {Jue}}, \ and\ \bibinfo {author}
  {\bibfnamefont {T.~J.}\ \bibnamefont {Silva}},\ }\href
  {http://dx.doi.org/10.1038/nphys3418} {\bibfield  {journal} {\bibinfo
  {journal} {Nat. Phys.}\ }\textbf {\bibinfo {volume} {11}},\ \bibinfo {pages}
  {825} (\bibinfo {year} {2015})}\BibitemShut {NoStop}%
\bibitem [{\citenamefont {Belmeguenai}\ \emph {et~al.}(2015)\citenamefont
  {Belmeguenai}, \citenamefont {Adam}, \citenamefont {Roussign\'e},
  \citenamefont {Eimer}, \citenamefont {Devolder}, \citenamefont {Kim},
  \citenamefont {Cherif}, \citenamefont {Stashkevich},\ and\ \citenamefont
  {Thiaville}}]{Belmeguenai15}%
  \BibitemOpen
  \bibfield  {author} {\bibinfo {author} {\bibfnamefont {M.}~\bibnamefont
  {Belmeguenai}}, \bibinfo {author} {\bibfnamefont {J.-P.}\ \bibnamefont
  {Adam}}, \bibinfo {author} {\bibfnamefont {Y.}~\bibnamefont {Roussign\'e}},
  \bibinfo {author} {\bibfnamefont {S.}~\bibnamefont {Eimer}}, \bibinfo
  {author} {\bibfnamefont {T.}~\bibnamefont {Devolder}}, \bibinfo {author}
  {\bibfnamefont {J.-V.}\ \bibnamefont {Kim}}, \bibinfo {author} {\bibfnamefont
  {S.~M.}\ \bibnamefont {Cherif}}, \bibinfo {author} {\bibfnamefont
  {A.}~\bibnamefont {Stashkevich}}, \ and\ \bibinfo {author} {\bibfnamefont
  {A.}~\bibnamefont {Thiaville}},\ }\href {\doibase 10.1103/PhysRevB.91.180405}
  {\bibfield  {journal} {\bibinfo  {journal} {Phys. Rev. B}\ }\textbf {\bibinfo
  {volume} {91}},\ \bibinfo {pages} {180405} (\bibinfo {year}
  {2015})}\BibitemShut {NoStop}%
\bibitem [{\citenamefont {Vansteenkiste}\ \emph {et~al.}(2014)\citenamefont
  {Vansteenkiste}, \citenamefont {Leliaert}, \citenamefont {Dvornik},
  \citenamefont {Helsen}, \citenamefont {Garcia-Sanchez},\ and\ \citenamefont
  {Van~Waeyenberge}}]{Vansteenkiste14}%
  \BibitemOpen
  \bibfield  {author} {\bibinfo {author} {\bibfnamefont {A.}~\bibnamefont
  {Vansteenkiste}}, \bibinfo {author} {\bibfnamefont {J.}~\bibnamefont
  {Leliaert}}, \bibinfo {author} {\bibfnamefont {M.}~\bibnamefont {Dvornik}},
  \bibinfo {author} {\bibfnamefont {M.}~\bibnamefont {Helsen}}, \bibinfo
  {author} {\bibfnamefont {F.}~\bibnamefont {Garcia-Sanchez}}, \ and\ \bibinfo
  {author} {\bibfnamefont {B.}~\bibnamefont {Van~Waeyenberge}},\ }\href
  {http://scitation.aip.org/content/aip/journal/adva/4/10/10.1063/1.4899186}
  {\bibfield  {journal} {\bibinfo  {journal} {AIP Adv.}\ }\textbf {\bibinfo
  {volume} {4}},\ \bibinfo {eid} {107133} (\bibinfo {year} {2014})}\BibitemShut
  {NoStop}%
\bibitem [{\citenamefont {Wagner}\ \emph {et~al.}(2015)\citenamefont {Wagner},
  \citenamefont {Stienen},\ and\ \citenamefont {Farle}}]{Wagner15}%
  \BibitemOpen
  \bibfield  {author} {\bibinfo {author} {\bibfnamefont {K.}~\bibnamefont
  {Wagner}}, \bibinfo {author} {\bibfnamefont {S.}~\bibnamefont {Stienen}}, \
  and\ \bibinfo {author} {\bibfnamefont {M.}~\bibnamefont {Farle}},\ }\href
  {https://arxiv.org/abs/1506.05292v1} {\bibfield  {journal} {\bibinfo
  {journal} {arXiv:1506.05292}\ } (\bibinfo {year} {2015})}\BibitemShut
  {NoStop}%
\end{thebibliography}%
\end{document}